\documentclass{article}
\usepackage[T1]{fontenc}
\usepackage{aecompl}
\usepackage{amsmath,amsfonts,amsthm}
\usepackage{mathtools}
\usepackage{commath}
\usepackage[sc,osf]{mathpazo}
\usepackage{hyperref}
\usepackage{stmaryrd}%double brackets
\usepackage{enumitem,amssymb}
\usepackage[dvipsnames]{xcolor}
\usepackage{graphicx,scalerel}%chapeau proportionel
\usepackage{float}
\usepackage{ucs}
\usepackage[utf8x]{inputenc}
\usepackage{cleveref}
\usepackage{url}
\usepackage{ulem}
\usepackage{caption}
\usepackage{subcaption}
\usepackage{authblk}
\usepackage{nicematrix}

\makeatletter
\def\blfootnote{\gdef\@thefnmark{}\@footnotetext}
\makeatother

\newcommand\nnfootnote[1]{%
  \begin{NoHyper}
  \renewcommand\thefootnote{}\footnote{#1}%
  \addtocounter{footnote}{-1}%
  \end{NoHyper}
}

\newlist{todolist}{itemize}{2}
\setlist[todolist]{label=$\square$}

\definecolor{dr}{RGB}{119,23,20}
\newtheoremstyle{break}
  {11pt}{11pt}%
  {\color{dr}\itshape}{}%
  {\color{dr}\bfseries}{}%
  {\newline}{}%
\theoremstyle{break}

\definecolor{dg}{RGB}{23,119,20}
\newtheoremstyle{breakg}
  {11pt}{11pt}%
  {\color{dg}\itshape}{}%
  {\color{dg}\bfseries}{}%
  {\newline}{}%
\theoremstyle{breakg}

\definecolor{db}{RGB}{23,20,119}
\newtheoremstyle{dotlessP}{}{}{\color{db}}{}{\color{db}\bfseries}{}{\newline }{}

\theoremstyle{dotlessP}

\usepackage{tcolorbox}

\tcbuselibrary{theorems}
\newtcbtheorem
  []% init options
  {definitionbox}% name
  {Definition}% title
  {%
    colback=red!5,
    colframe=red!65!black,
    coltitle=black,
    fonttitle=\bfseries,
  }% options
  {def}% prefix

\newtcbtheorem
  []% init options
  {theorembox}% name
  {Theorem}% title
  {%
    colback=red!1,
    colframe=red!20!black,
    coltitle=red,
    fonttitle=\bfseries,
  }% options
  {thm}% prefix

\newcommand{\Cov}{\mathrm{Cov}}

\newcommand{\Fm}{{\bf F}}

\newcommand{\Um}{{\bf U}}

\newcommand{\Xm}{{\bf X}}

\newcommand{\Lambdam}{\boldsymbol{\Lambda}}

\newsavebox{\tempbox}

\title{Impact of the COVID-19 pandemic on bike-sharing uses in two french towns  }

\author[1]{Angelo Furno}
\affil[1]{LICIT-ECO7, UMR\_T9401, ENTPE-Gustave Eiffel University, ENTPE, Cit\'e des Mobilit\'es, 69675 Bron, France.}
\author[2]{Bertrand Jouve$^*$}
\affil[2]{LISST, UMR5193, Toulouse Jean Jaur\`es University, CNRS, 5 all\'ees A. Machado, 31058 Toulouse Cedex 9, France}
\author[2]{Bruno Revelli}
\author[3]{Paul Rochet}
\affil[3]{DEVI ENAC, 7 avenue E. Belin, 31055 Toulouse Cedex 4, Toulouse, France}
\author[2]{Alix Rigal}
\author[2]{Najla Touati}

\date{Juin 2021}

\begin{document}
\maketitle

\abstract{Urban areas have been dramatically impacted by the sudden and fast spread of the COVID-19 pandemic. As one of the most noticeable consequences of the pandemic, people have quickly reconsidered their travel options to minimize infection risk. Many studies on the Bike Sharing System (BSS) of several towns have shown that, in this context, cycling appears as a resilient, safe and very reliable mobility option. Differences and similarities exist about how people reacted depending on the place being considered, and it is paramount to identify and understand such reactions in the aftermath of an event in order to successfully foster permanent changes. In this paper, we carry out a comparative analysis of the effects of the pandemic on BSS usage in two French towns, Toulouse and Lyon. We used Origin/Destination data for the two years 2019 (pre-pandemic) and 2020 (pandemic), and considered two complementary quantitative approaches. Our results confirm that cycling increased during the pandemic, more significantly in Lyon than in Toulouse, with rush times remaining exactly the same as during the pre-pandemic year. Among several results, we note for example that BSS usage is more evenly spread throughout the day in 2020, peripheral/city center flow is more noticeable in Toulouse than in Lyon and that student BSS usage is more specific in Lyon. We also found that trip duration during the pandemic situation was longer on working days and shorter on weekends.}

\section{Introduction}
\nnfootnote{$\dagger$ This research is party funded by the ANR projects ANR-21-COVR-0027-01 and the LABEX SMS ANR-11-LABEX-0066.}
\nnfootnote{$*$ Corresponding author: bertrand.jouve@univ-tlse2.fr. The authors are sorted by alphabetic order.}
The rapid spread of the COVID-19 pandemic has brought unprecedented challenges and has been threatening normal human life and global public health. Since the first identified case of COVID-19 in December 2019 in the city of Wuhan in China, measures have been taken worldwide and at different scales to stop virus spread. They concern of course the field of health (research and administration of vaccines, rapid identification of new clusters, etc.), but also transportation (limitation of international flights, temporary bans on inter-regional transportation, etc.) as human mobility has significantly contributed to virus propagation. In urban areas, many people voluntarily chose to avoid using public transport, especially due to difficulties to comply with barrier measures when traveling. 
In this context, individuals often re-evaluated their travel options towards more isolated modes such as private cars, personal or shared bicycles \cite{bert2020, bucsky2020}. 

Considerations of health risks, travel flexibility, traffic reduction, or the desire to spend less money in times of economic crisis now shape individual decisions about transportation. 
More than two years after the beginning of the pandemic, the eradication of COVID-19 and its variants remains uncertain in the short term \cite{barbieri2021} and the emergence of new pandemics is a scenario now considered highly probable by the scientific community. Under these circumstances, several observations show that cycling can be a valuable choice \cite {pase2020, fuller2019, saberi2018, song2022}, and, according to some, even contribute to faster economic recovery. In many cities, public authorities are implementing temporary facilities, strategies derived from ``tactical'' urban planning, to accompany this increase in cycling \cite{chassignet2020, silva2016}. Cycling is a largely flexible, inexpensive mode of urban transport easily adaptable to a wide range of situations, compared to, for example, public transport, whose adaptation is often a longer-term challenge \cite{gkiotsalitis2021}.

The effects of the pandemic on urban mobility have emphasized and amplified some of the gradual transformations underway so far as a result of the awareness of the ecological emergency \cite{AR6_IPCC_WIII}. These transformations tend to favor public transport and active or ``soft'' mobilities \cite{beatley2014, rojas2021}. 
Among these, cycling, as opposed to public transport, encourages physical activity, provides greater freedom of use, reduces environmental impact \cite{zhang2021} and only requires relatively low-cost deployment. On the other hand, its use is largely dependent on the weather, topography and urban environment (speed moderation and bicycle facilities \cite{bergantino2021}).  The development of cycling has been encouraged by public authorities by setting up Bike Sharing Systems (BSS), appropriate infrastructures and, more recently and in some countries, by financial incentives for the purchase or repair of bicycles. These policies are part of a new paradigm which advocates sustainable, multimodal mobility (as opposed to the single-mode, infrastructure-focused paradigms of the automobile and public transport). Thanks to their flexibility in multimodal trips, BSS uses are strongly related to rail and tube connectivity \cite{bocker2020} in a noticeably different manner in urban centers and suburban areas \cite{wu2021, liu2020}. However, whenever possible, door-to-door trips are favored \cite{frere2015}. 

Cycling thus appears to be a resilient and reliable urban mobility option in the short and long term, compatible with both the health crisis as we know it and the objectives of sustainable development. Since mobility habits are difficult to alter \cite{rocci2015, moro2018, buchel2022}, taking advantage of changes in behaviors during the current health crisis and of a cognitive appropriation of cycling sometimes facilitated by the economic situation is an opportunity and a challenge for many cities to sustainably transform urban mobility practices and habits. 

The development of BSS is an effective response by public authorities to increase bicycle use in urban areas \cite{demaio2009, shaheen2012, heinen2010c, pucher2010}. For example, in the year following the introduction of Velo'v (the BSS deployed in the city of Lyon), a 44\% increase in bicycle trips \cite{buhrmann2007} was observed, with 96\% new users who had not previously cycled in Lyon's city center. Similarly, in Paris, cycling increased by 70\% with the launch of the Vélib BSS \cite{nadal2008velib, shaheen2010}. While effective in the short term, several studies have shown that the development of bike sharing also contributes to a sustainable increase in the cycling population \cite{demaio2009} and that BSS usage is related to bike usage as a whole \cite{pazdan2021}. Today, a little more than half a century after the first bike-sharing system was implemented in 1965 in Amsterdam, BSS have been developed in many cities around the world and grown increasingly successful \cite{shaheen2010, ma2020, eren2020} while often relying on quite similar technological solutions. BSS appeared in France with a pioneer initiative in La Rochelle in 1976 \cite{Hure2016} and a fully automated solution (probably the first in the world) was implemented in Rennes in 1998. But, in France, the generation of BSS as we have it today (with the need for each user to register a personal account) was first introduced in Lyon in 2005, followed by Paris and Toulouse in 2007. 

The success of bike-sharing systems all over the world has given rise to a rapid increase in the scientific research on these systems. There have been several literature review papers summarizing the research done in this field \cite {eren2020, teixeira2021, fishman2019, medard2021}.

The effect of the pandemic on bike share usage has been widely studied, with data mostly from the year 2020 and collected from cities all over the world. Some studies use operator data, either Origin/Destination or simply traffic data. Among these are studies comparing five cities in the USA \cite{tokey2020}, about New-York \cite {teixeira2020, wang2021, pase2020}, Chicago \cite{hu2021}, Lisbon \cite{teixeira2021}, London \cite{li2021}, Singapore \cite{song2021}, Zurich \cite{li2020}, Beijing \cite {chai2020, shang2021}, Kosice Slovakia \cite{kubalak2021}. Several other studies are based on online or on-site surveys in Thessaloniki in Greece \cite{niki2020}, San Antonio in the USA \cite{jobe2021}, in several Italian towns \cite{bergantino2021}, distributed in several European towns \cite{monterde2022}, or with participants from all over the world \cite {bert2020, barbieri2021, chan2020}. 

In a recent paper, \cite{song2022} summarizes recent progress in bike-share studies related to COVID-19. They classify the issues in three main domains. The first one focuses on spatio-temporal changing patterns in bike-share usage before and after the pandemic crisis, mainly with data analysis methods applied on BSS data. The second one is more about comparing different modes of transport. The third one is centered on user behaviors via survey analysis. \cite{heydari2021} provides an interesting summary of the diversity of published results. 

Let us focus on some relevant recent papers falling within the first type of approach. 

With a Bayesian time-series model on monthly aggregated data, \cite{heydari2021} estimates the impact of the pandemic on both trip duration and number of trips in London. Confirming previous results, in (e.g., \cite{li2021,wang2021}), the authors observe that trips are longer than expected during and after the spring 2020 lockdown.  

\cite{song2022} uses an original graph theoretical approach for comparing flows of bikes during 4 COVID-relevant periods in Singapore. The proposed approach provides an adequate means to measure the increase in local trips observed during and after the lockdown. \cite{pase2020} also uses a graph theoretical approach on New-York BSS data showing the reconfiguration of BSS trips on Manhattan Island during the lockdown. 

\cite{tokey2020} leverages a General Linear Model (GLM) to formalize the association of the month-to-month change in bike-share activity within a geographical tract and its distance from the CBD. Using weekly aggregated data for 6 towns in the USA over 6 months, they confirm previous results that distance to the CBD is positively correlated to BSS usage \cite {faghih2014, wang2016, noland2016, yang2020} and clearly explains the dynamics of change due to COVID-19: CBD areas are more responsive to the pandemic with usage declining faster but also recovering faster. 

To precisely evaluate the impact of the lockdown (and lockdown ease) on BSS usage in London, \cite{li2021} crosses two models: a segmented regression model and a Bayesian structural time-series model, both on aggregated data at daily level for the period from January 2019 to June 2020. Rainy conditions and BSS 2019 trips are included in the regression model as binary covariates. The authors distinguish morning and evening peaks from the rest of the day and three levels of travel duration (short, middle, long). Their main findings are that the number of long trips increased during the lockdown and lockdown ease, which was not observed for short trips, and that the lockdown ease had little effect on the morning peak which remained low throughout the lockdown. They also found that immediate effects on BSS usage were more important in highly infected boroughs than in low infected ones.  

With a similar goal of analyzing traffic-bike changes during the lockdown, \cite{buchel2022} uses data collected through 13 and 20 bike counters respectively in Basel and Zurich (Swiss). To estimate daily cycling traffic, two Random Forest Regressors are used: one for Working days and one for Weekend days. Both models depend on the weather, time of the day and whether the day is a holiday (school or public). Among the main findings, daily traffic during the lockdown is noted as more complicated than just a mix between pre-lockdown working and weekend days traffic patterns. It is also found that bike traffic during the second COVID wave followed the same pattern observed during the first one. After the first lockdown, bike usage increased more in Zurich than in Basel although cycling culture is more widely developed in Basel than in Zurich.

The study reported in \cite{hu2021} is very similar to ours. Instead of a log-linear regression model, the authors used Generalized Additive Models (GAMs) for comparing regular BSS usage between March and July 2019 in Chicago with the pandemic BSS usage for the same period in 2020. They considered weather conditions, seasonality and holidays effects. Like previous studies, they observed that trips were longer in 2020 than in 2019, and bike-sharing more resilient than other transport modes. They also showed that the pandemic impact on bike sharing was differential, depending on socio-demographic patterns around each dock-station.

\cite{shang2021} proposes a novel method to calculate bikers' trajectories and estimate the environmental benefits during COVID-19. The exact route taken by cyclists could prove to be rather complex to infer. With the same method as the one proposed in \cite{shang2021} for Beijing data, we used OSMnx and OpenStreetMap to calculate the shortest paths between origin and destination for Toulouse BSS data. Checking the ``bike'' option in OpenStreetMap proved to be a poor solution, leading to many highly improbable speeds. Checking the ``pedestrian'' option proved instead to be a much better approach. It seems to indicate that Toulouse BSS riders take shortcuts, not necessarily referenced as cycle routes. In a still ongoing survey in Toulouse, we asked 125 riders to describe their route and the results show that 88\% match the ``pedestrian'' option of OSMnx but the remaining 12\% do not follow clear logics. Because of this difficulty, in the following, we will not use any hypotheses on routes.

In this paper, we propose a spatio-temporal analysis of the COVID-19 impact on BSS data usage in two French towns, Toulouse and Lyon, which were among the first cities in the world to implement automated BSS and, therefore, where BSS use is well established. For this purpose, we consider 2019 as a baseline year against which we can compare corresponding periods in 2020. To our knowledge, this is the first study of this type on French cities. In contrast to many previous papers, we use several levels of data aggregation (week, day, 10 minutes) to improve the accuracy of the results. After describing the data (Section \ref{data}), the following section (Section \ref{time-modeling}) provides general statistics and shows that the number of bikes leaving or returning a dock-station is accurately modeled by a log-linear function of time variables and of the rain level. Section \ref {sec:efa} uses matrix factorization to highlight the presence of latent explanatory factors, which, for instance, can depend on the spatial distribution of the stations on the territory, and study their dynamics before and during the COVID pandemic. Finally, Section \ref {Conclusion} provides some interpretations of the results comparing the two methods and data from Public Transport and road traffic.

\noindent Calculations were made using R software and python language, and maps were produced with QGIS software.

\section{Data description and context}
\label{data}

\begin{figure}[H]
\centering
\includegraphics[width=0.9\linewidth]{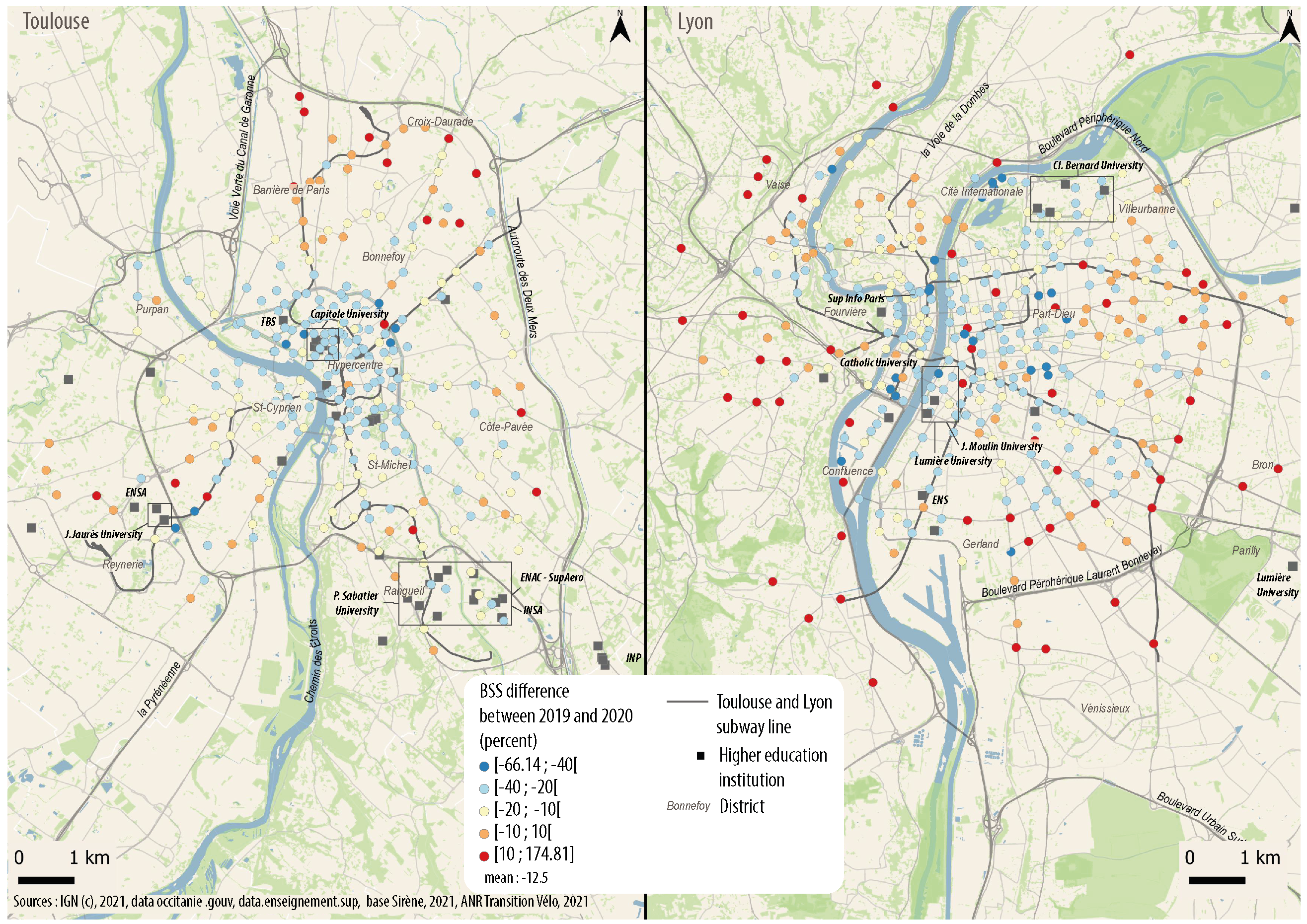}
\caption{Locations of bike-sharing docks in Toulouse and Lyon. Each station is colored according to the percentage of traffic increase/decrease from 2019 to 2020. To avoid bias in case a station is out of order for a given period, the corresponding periods of both 2019 and 2020 were dismissed in the calculation of the percentage.}
\label{fig:localisation}
\end{figure}

In Lyon and Toulouse, a dock-based Bike-Sharing System (BSS) was installed by JC Decaux in 2005 and 2007 respectively. Since \textsuperscript{th} February 20 2020, and progressively over the year, half of the Lyon fleet has been equipped with electric assistance which riders may choose to use or not. From our data, we have not been able to differentiate both types of users but we did not observe any significant increase in usage when electric assistance was implemented. So, we used the collection of all the Origin/Destination trips relative to JC Decaux BSS for both towns and for the two years 2019 and 2020, without distinguishing electric or non-electric usage. Each row of the dataset includes departure time and arrival time (with an accuracy to the minute), origin station and destination station (with geographical coordinates). The year 2019 will be considered as a ``normal'' year, as opposed to 2020 when France faced the onslaught of the first wave of COVID-19. For each year and each town, a record of the dataset contains the positioning and timing information of locking and unlocking of bikes, excluding that of rebalancing operations. Because data are anonymized, it is not possible to follow a user in his/her daily use of BSS.
We removed trips with duration less than or equal to 2 min with identical origin and destination, or with a duration more than 12 hours regardless of origin or destination. These data accounted for about 2.3\% of each dataset, except for Lyon 2019 for which it is 3.7\% with half of them being trips with identical origin and destination and of less than 1 minute. Please note that our data correspond to real trips, that is, service operations are counted apart.
Following these filters, we considered 287 dock-stations for Toulouse (no change between 2019 and 2020), for 3.83 and 2.95 million trips in respectively 2019 and 2020 (23\% decrease). Lyon had 408 such dock-stations in 2019 and 430 in 2020, for respectively 8.11 and 6.96 million trips (14\% decrease). As usual, the density of dock-stations is higher in city centers (\Cref{fig:localisation}). Total population is about 1.37 higher in Lyon-Villeurbanne than in Toulouse (about 670000 people against 490000) for a number of BSS trips more than twice as high in Lyon. It reveals that the BSS network is better established in Lyon than in Toulouse, which is also revealed by the mean number of trips per bike which is about 50\% higher in Lyon. 

Traffic changes between 2019 and 2020 are not homogeneous among the dock-stations: while city centers show a high decrease in uses, more periphery and residential dock-stations are less impacted and can even show some increasing usage (\Cref{fig:localisation}). These observations are consistent with
\cite{chai2020, hu2021, kim2021} observations in Beijing, Chicago and Seoul respectively, but inconsistent with \cite{hua2021} results in Nanjing (China).

Dockless systems have emerged, but they are a complementary service to docked-BSS, with significantly different impact and users \cite{ma2020}, and will not be discussed in this paper. 

We also have Public-Transport and traffic data for both periods and both towns. The aim of this paper is not so much to specifically analyze these data as to give an overview of the various transport dynamics in both towns and make it easier to interpret the result on BSS use dynamics. We used Floating Car Data (FCD) to evaluate traffic dynamics. The principle of FCD is to collect real-time traffic data by locating vehicles via mobile phones or GPS over the entire road network \cite{altintasi2017}. As for Public Transport data, we have the number of validations per 15 minutes for the different networks (Bus, Tramway and Subway). Each validation corresponds to an entry of one ID user somewhere in the transport network. Users are anonymized in such a way that two entries of the same user have different IDs and then users ‘daily-usage is not reconstructed.  

Weather observation data were provided by Météo-France for each hour of both years. For each town, we considered data from the station closest to the city-center (Bron and Blagnac for respectively Lyon and Toulouse). The "rain level" variable has the highest negative correlation with BSS usages, which is consistent with previous findings \cite{bean2021}. Moreover, we observed strong correlations between several weather-related variables. Therefore, in the following we will just consider "rain level". 

Both cities have similar dynamics in pandemic development with rates of intensive care hospitalization following the same pattern \footnote{https://www.data.gouv.fr/fr/datasets/donnees-hospitalieres-relatives-a-lepidemie-de-covid-19/}. After a peak at the end of March 2020, these rates were very low during the 3 summer months (June-July-August) and then rose again at the end of September to reach the peak of the second wave around November 10, 2020. Because of these similarities, data relative to COVID cases will not be taken into account in our analysis.     
Finally, below is a short timeline of the main 2020 dates of pandemic-related measures taken in France:
\begin{itemize}
    \item January 24: the French Minister of Solidarity and Health confirmed the first cases of COVID-19 in France;
    \item March 12 and 13: closure of nurseries, schools, colleges, high schools and universities; extension of the winter break for two months. Gatherings of more than 100 people were prohibited;
    \item March 17: beginning of the first lockdown;
    \item May 11: end of the lockdown, but meetings with more than 10 persons remained forbidden;
    \item June 22: reopening of schools;
    \item October 17: curfew from 9PM to 6AM declared in the largest cities, including Toulouse and Lyon;
    \item October 30: beginning of the second lockdown. Non-essential stores were closed down and circulation limited, online courses became mandatory for universities but schools remained open. 
    \item November 28: beginning of ease lockdown, with less restrictions on the circulation and reopening of stores;
    \item December 15: end of lockdown but curfew from 9PM to 6AM declared all over the national territory.
\end{itemize}

\section{Time modeling for citywide bike sharing use}
\label{time-modeling}
For each city, BSS use is measured every 10 minutes over the course of the years 2019 and 2020 yielding a sample of size $365 \times 144 = 52560$ for 2019 and $52704$ for the leap year 2020. The data consist of the raw number of trips over the whole city initiated in each 10-minute segment, without any pre-processing. \\

Temporal bike usage is modeled as a multiplicative function of time and amount of rain fall. The time component accounts for the time of the day (among the 144 10 min segments per day), the type of day (working or weekend day), the week of the year (1 to 53) and holidays. Rain variable is divided into three levels according to the duration of rain fall within the hour: Low (less than 20 min), Medium (between 21 and 40 min) and High (more than 41 min). \\

\noindent The proposed model, in the spirit of \cite{hu2021},
is as follows
\begin{equation}\label{model_temp} 
BSS \approx f(\text{time, type of day}) \times g(\text{week}) \times h(\text{holidays}) \times \ell(\text{precipitation}).
\end{equation}

The functions $f,g,h$ and $\ell$ are estimated non-parametrically via a linear model on the logarithm of $BSS$, with all variables treated as factors. Since the data are measured every 10 minutes, the estimation of the first function $f$ accounts for 144 parameters associated to the typical BSS use distribution during a working day (one for every 10 minute segment) and 144 parameters for the distribution over a weekend day. The remaining parameters correspond to the weekly evolution along the year (53 values) represented by the function $g$, an additional parameter adjusting for the 11 holidays of the year and 3 parameters for the three levels of rain fall. Accounting for identifiability, we reach a total of $343$ parameters for $52217$ (resp. $52361$) degrees of freedom for 2019 (resp. 2020) in both Toulouse and Lyon. The $R^2$ in the linear model varies between $0.97$ and $0.99$ in the four scenarios (Toulouse and Lyon in 2019 and 2020). Nearly all parameters are highly significant (p-value < $10^{-16}$) with at most three exceptions corresponding to night values with very low use. The amount of rain fall is the only non-temporal variable and, while it is highly significant in the model, it only accounts for less than $0.5\%$ of explained variance in each case.

Let us first notice that holidays induces a mean decrease of $25.65\%$ in BSS use. Similarly, the weather is responsible for a $23.1\%$ decrease in BSS use for a medium level of rain fall and for a $26.2\%$ in case of a high level, the two coefficients being highly significant. Nevertheless, the overall periodic time components (time of the day, type of day and week) accounts for more than $96 \%$ of explained variance, compared to less than $1 \%$ for holidays and rain fall.  \\

\begin{figure} 
\centering
\includegraphics[width=0.45\linewidth]{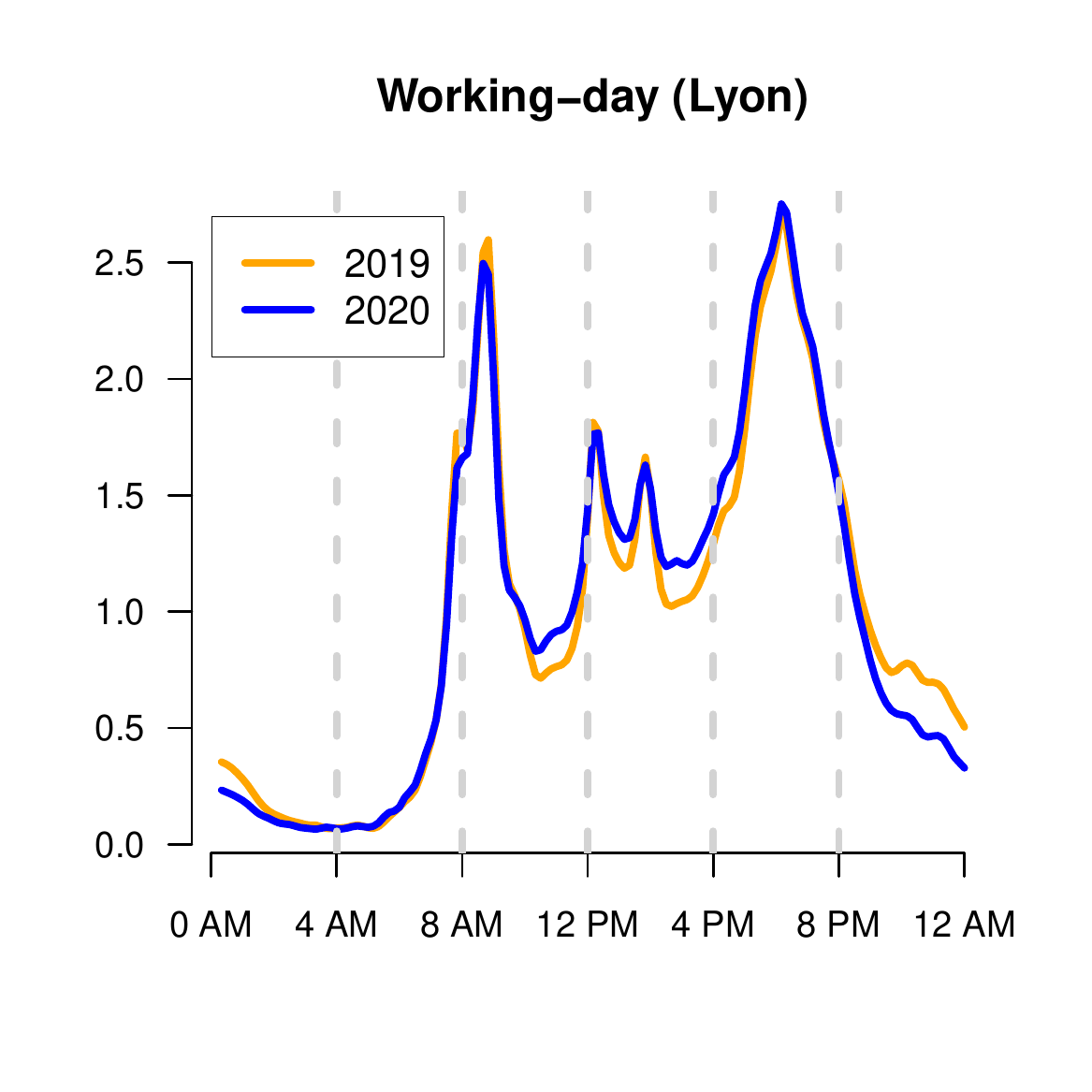}\null
\includegraphics[width=0.45\linewidth]{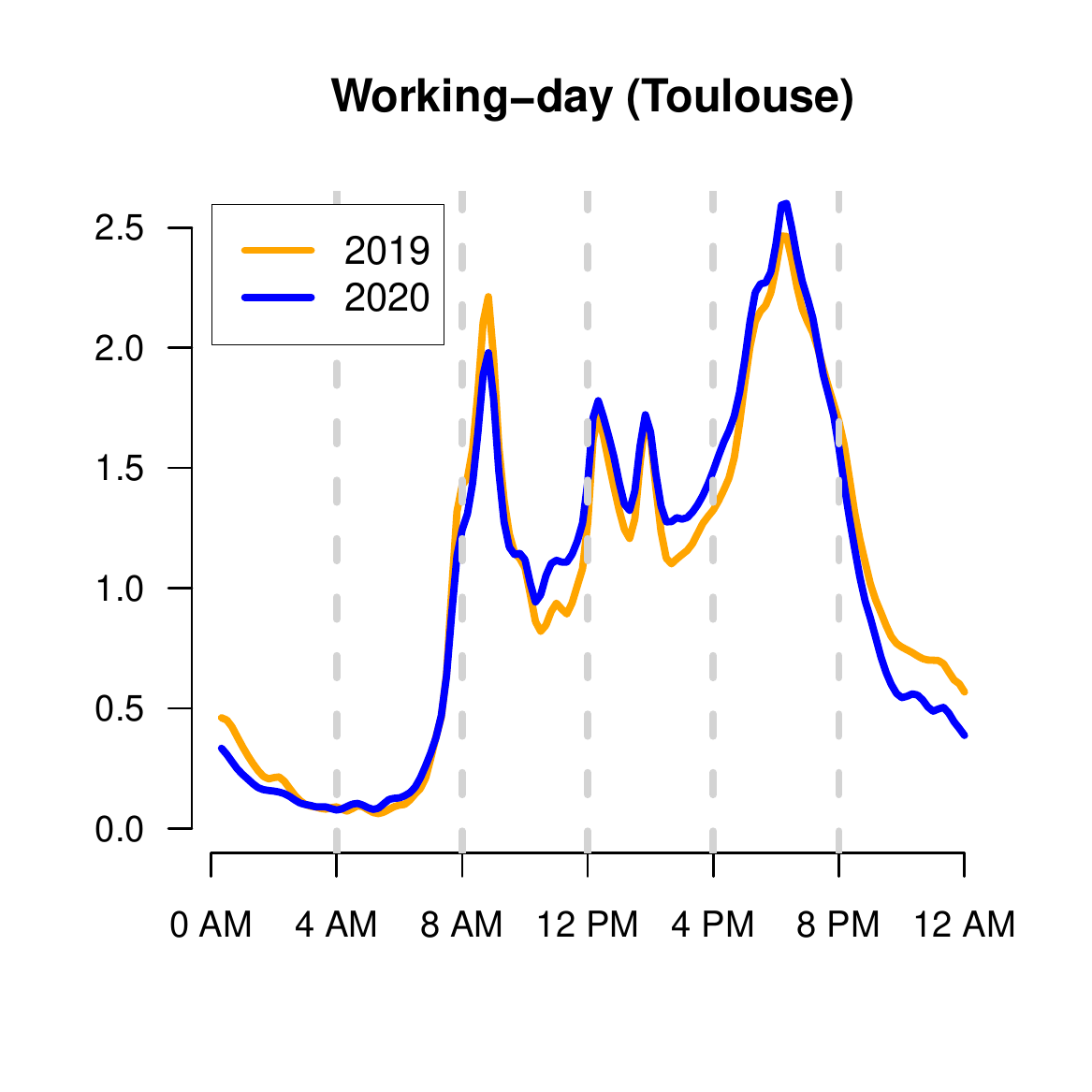}
\includegraphics[width=0.45\linewidth]{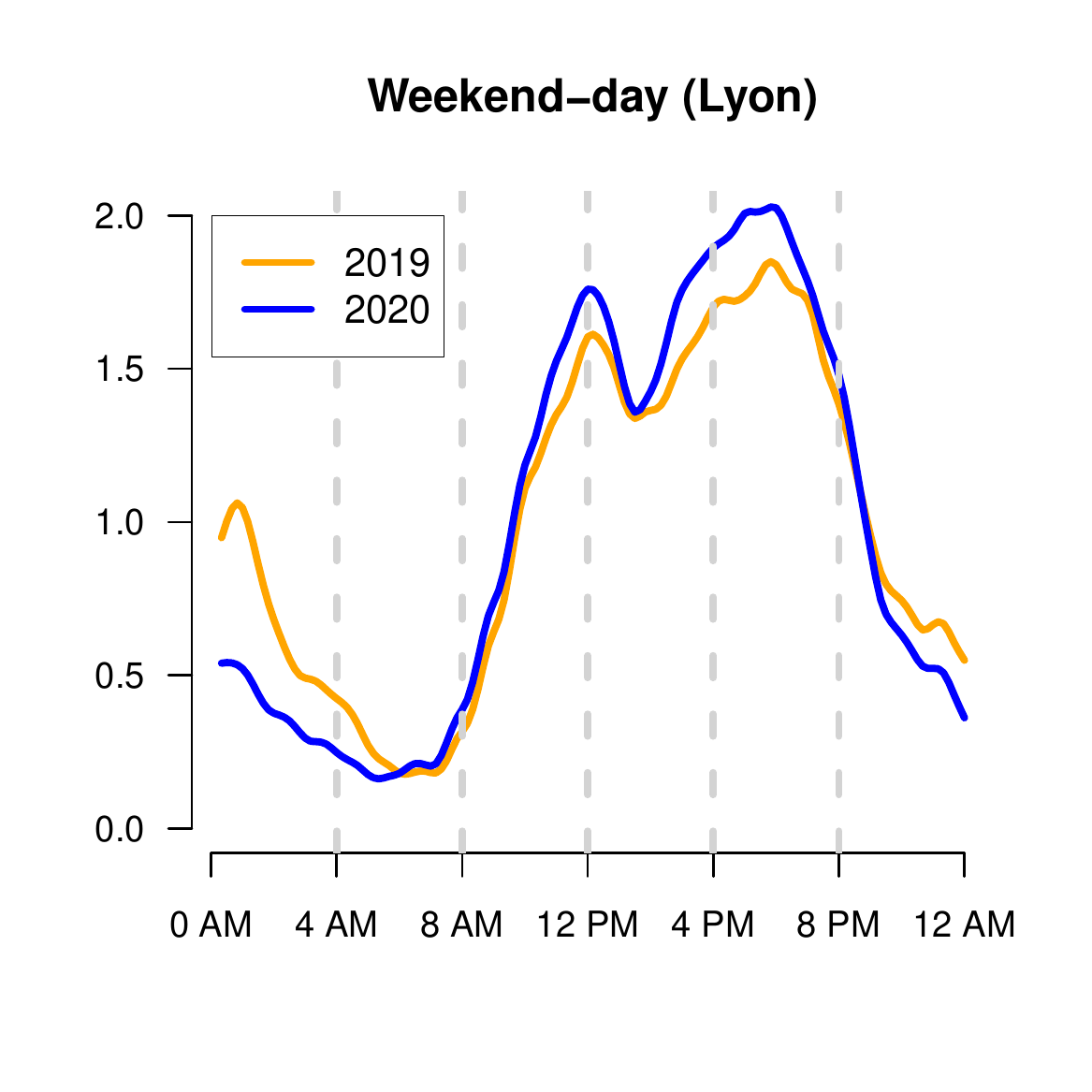}\null
\includegraphics[width=0.45\linewidth]{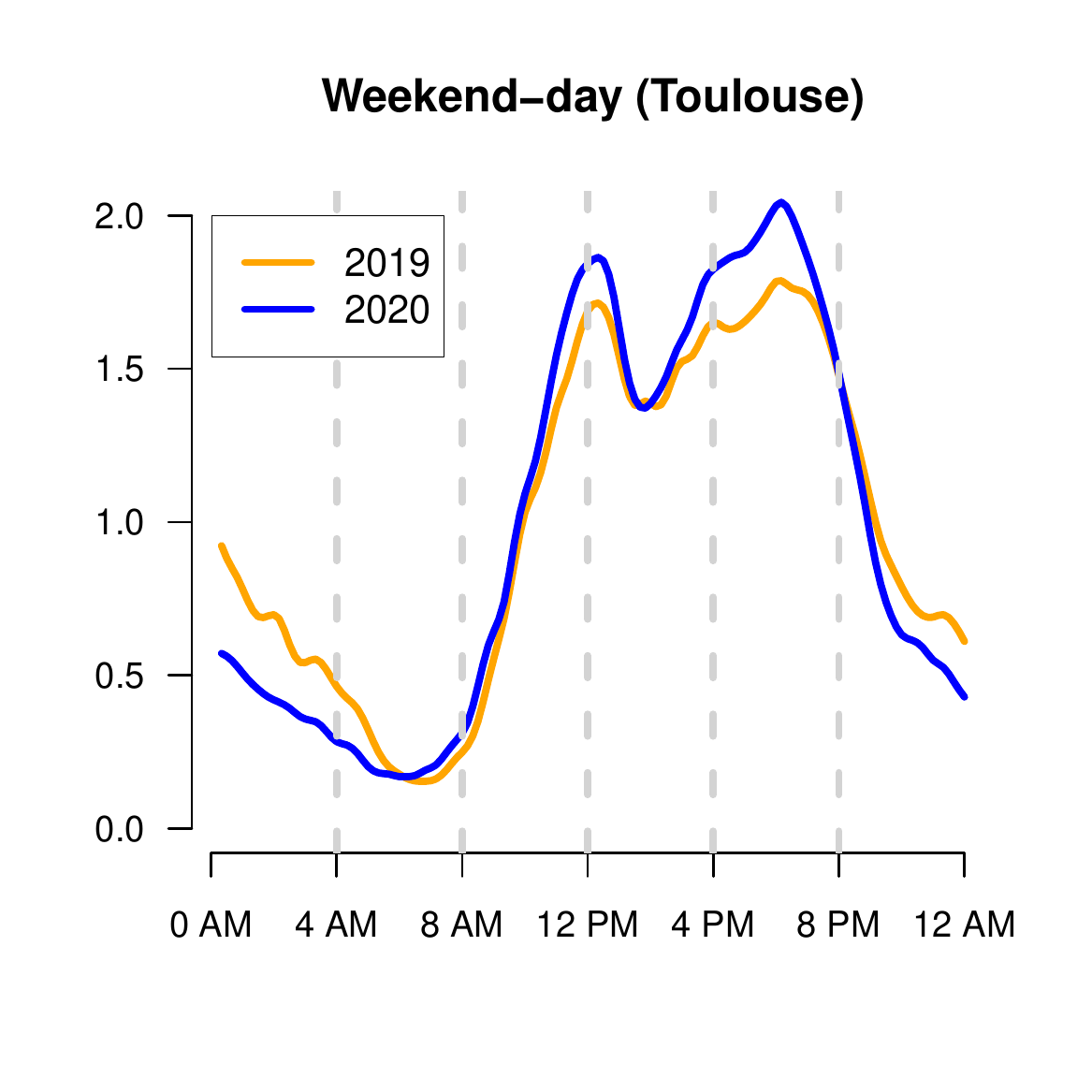}

\caption{Mean BSS use distribution over the course of a working-day and weekend-day in Lyon and Toulouse in 2019 and 2020 provided by the model. Curves are normalized so that the total sum of use over the year is equal to 1 for each curve. }
\label{fig:work_day}
\end{figure}

\begin{figure}[H]
\centering
\begin{subfigure}[b]{0.9\textwidth}
         \centering
         \includegraphics[width=1\linewidth]{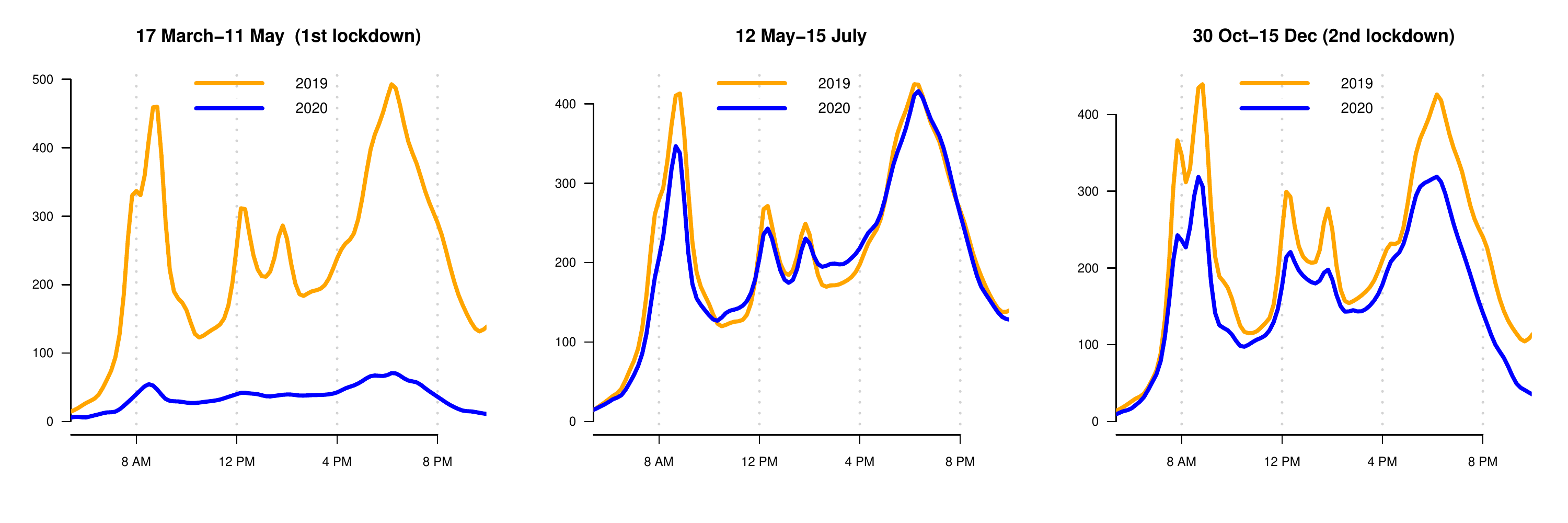}
         \caption{Lyon}
         \label{fig:Lyon}
     \end{subfigure}

\begin{subfigure}[b]{0.9\textwidth}
         \centering
         \includegraphics[width=1\linewidth]{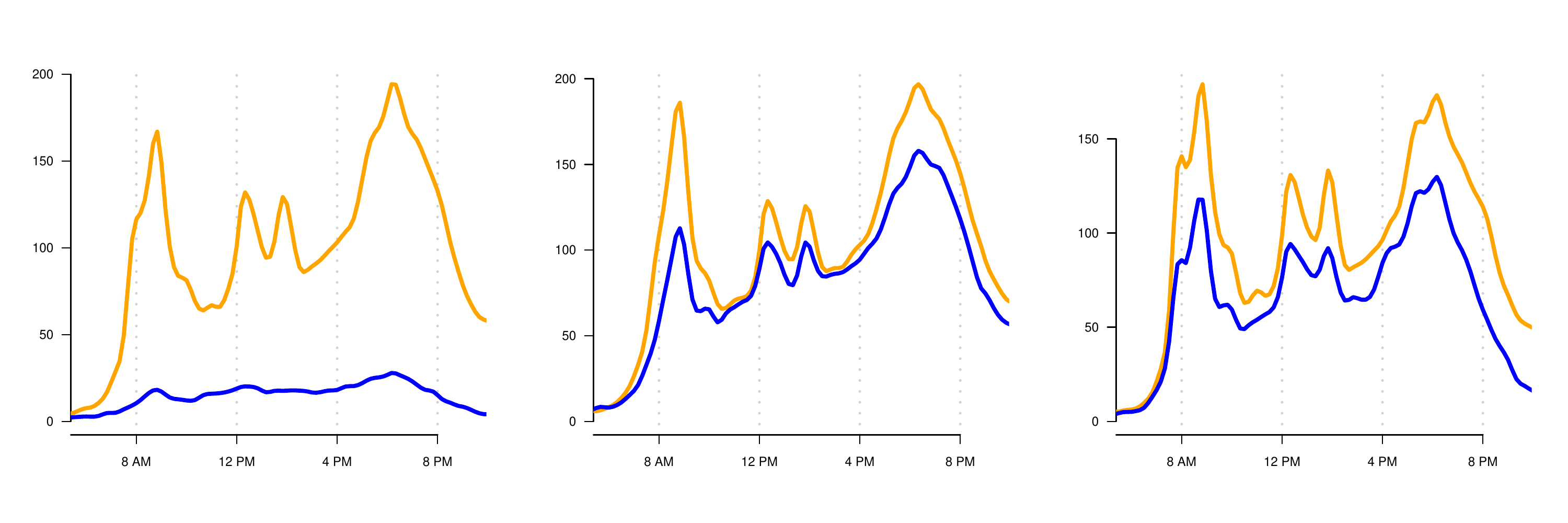}
         \caption{Toulouse}
         \label{fig:Toulouse}
     \end{subfigure}

\caption{Comparison of the average BSS use during a working-day in Lyon and Toulouse for the two lockdown periods and the period right after the first lockdown.}
\label{fig:comparison}
\end{figure}

\Cref{fig:work_day} are the daily components for our 4 models. Since the $R^2$ is nearly 1 for all 4 models, these curves can be considered as representative of the patterns of daily-BSS use corrected by weather and holidays biases. Because there is some variability depending on the period of the year (see \Cref{fig:comparison}), we did not normalize these curves by the number of annual trips and only the shapes of the curves should be considered. The distribution of daily BSS uses is extremely similar in 2019 and 2020 for working-days with a correlation of 0.989 in Lyon and 0.985 in Toulouse. The four visible traffic peaks occur at the same time in Toulouse and Lyon both in 2019 and 2020, namely around 8:45 AM, 12:15 PM, 1:45 PM and 6:15 PM. The cities show similar differences in behavior in 2020, with a small decrease in the 8:45 AM peak compensated by a higher and more spread-out use during late morning and the beginning of the afternoon. 
In contrast, the last three peaks almost perfectly overlap in 2019 and 2020. The decrease of BSS-use in the morning rush during the pandemic situation was observed in previous studies but often more pronounced \cite{hu2021,buchel2022,kubalak2021,li2020}. The morning peak completely disappeared during the pandemic situation, which is not what we observed even when focusing on different relevant periods of the year (\Cref{fig:comparison}). The relative increase in BSS use we observed around noon was also observed in some of these papers but a normalization of the data would be necessary to highlight them. What is surprising, however, is the absence of two peaks around noon in all previous studies. It may reflect different habits, or just the fact that these studies consider data with a one-hour accuracy, whose effect is to aggregate the two peaks. The use at night greatly decreased in 2020 in both Lyon and Toulouse, which can easily be explained by the curfews imposed by the French government. \\

On weekend-days, rhythms in Lyon and Toulouse are very similar with more variability in 2020 between the two peaks and a drop in usage around 1:30 PM. The presence of two peaks was observed in \cite{schimohr2021} but not in \cite{li2020}, \cite{gong2019} or \cite{buchel2022} where continuous increase in usage from noon to the end of the afternoon is probably due to data aggregation. The afternoon peak occurs slightly later compared to the highest congestion hours in these cities, situated around 5:45 PM.

We also observe a peak of traffic around 0:30AM in 2019, on weekends, especially for Lyon. In our Public Transport data for Lyon, it coincides with a peak of underground exits (preceded by a high number of entries just before and around midnight). It illustrates the importance of BSS/underground inter-modality, which is mainly asymmetric (BSS followed by the underground, or the other way round depending on the time of the day).

\begin{figure}[H]
\centering
\includegraphics[width=1\linewidth]{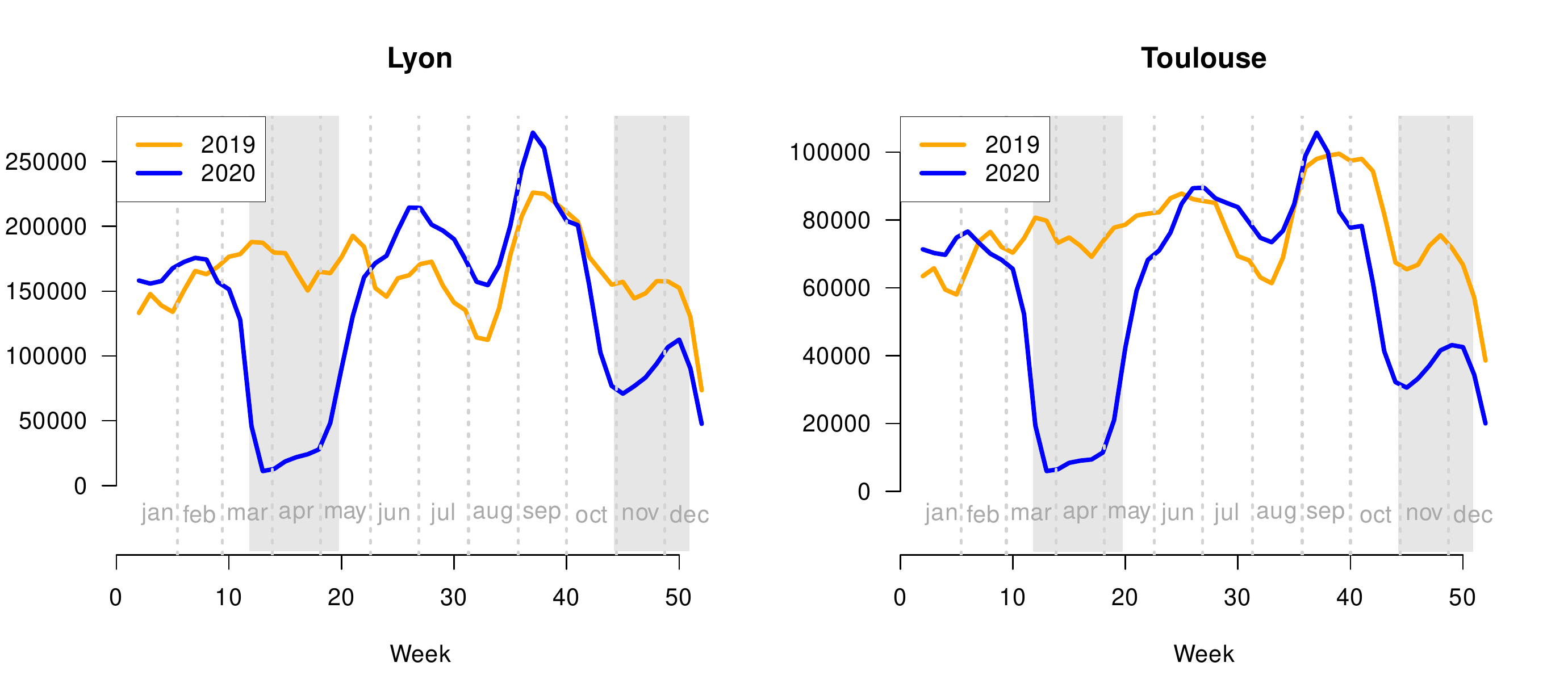}
\includegraphics[width=0.24\linewidth]{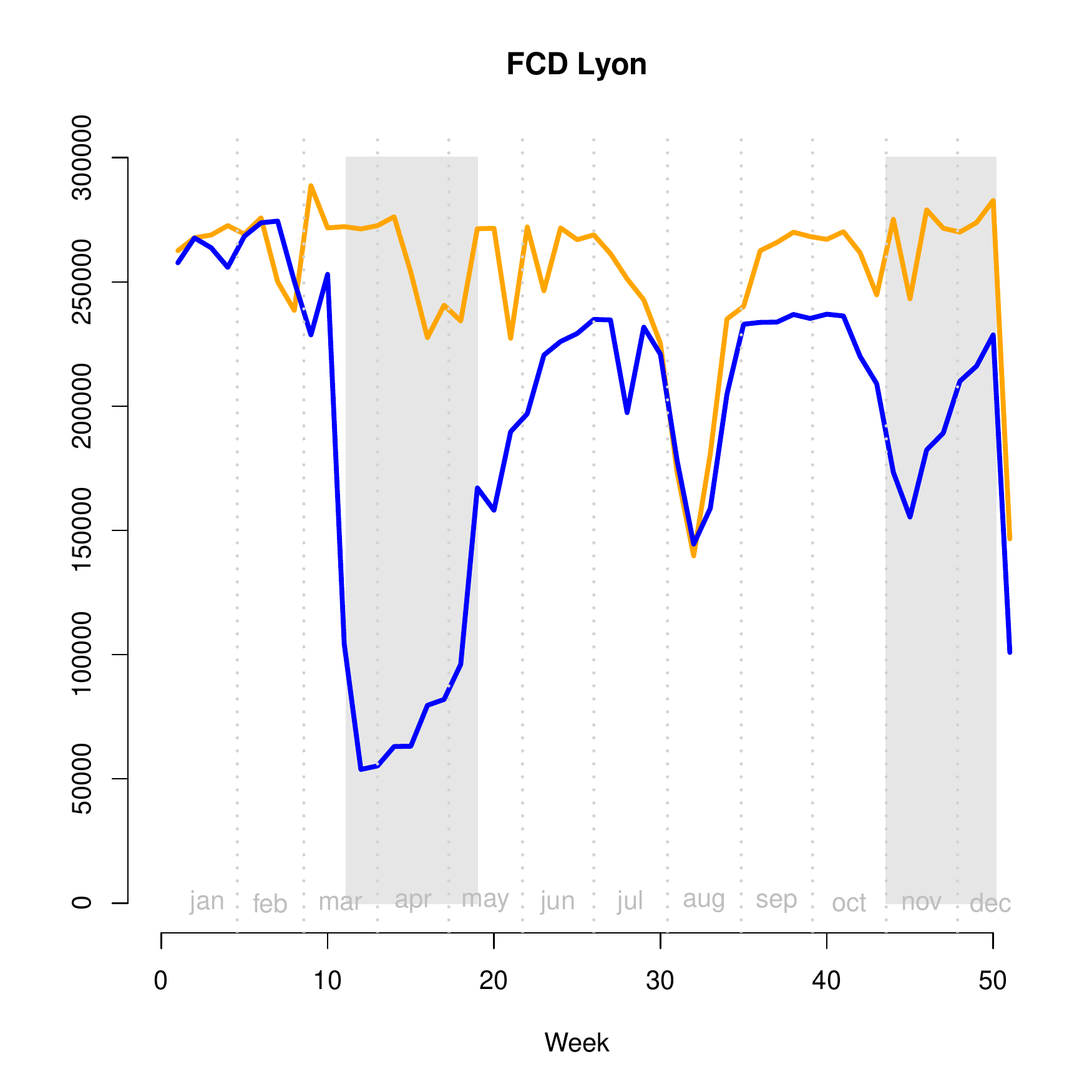}
\includegraphics[width=0.24\linewidth]{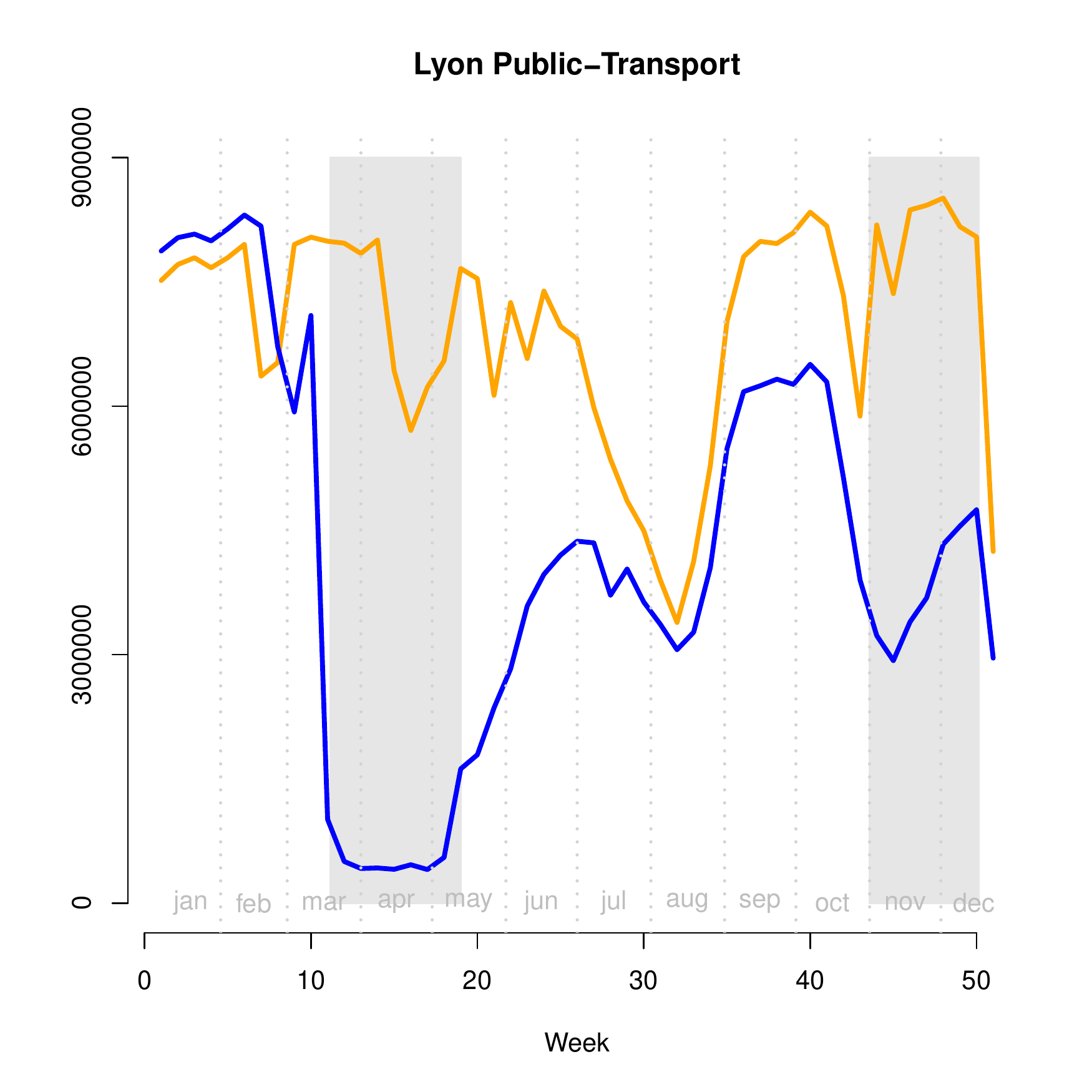}
\includegraphics[width=0.24\linewidth]{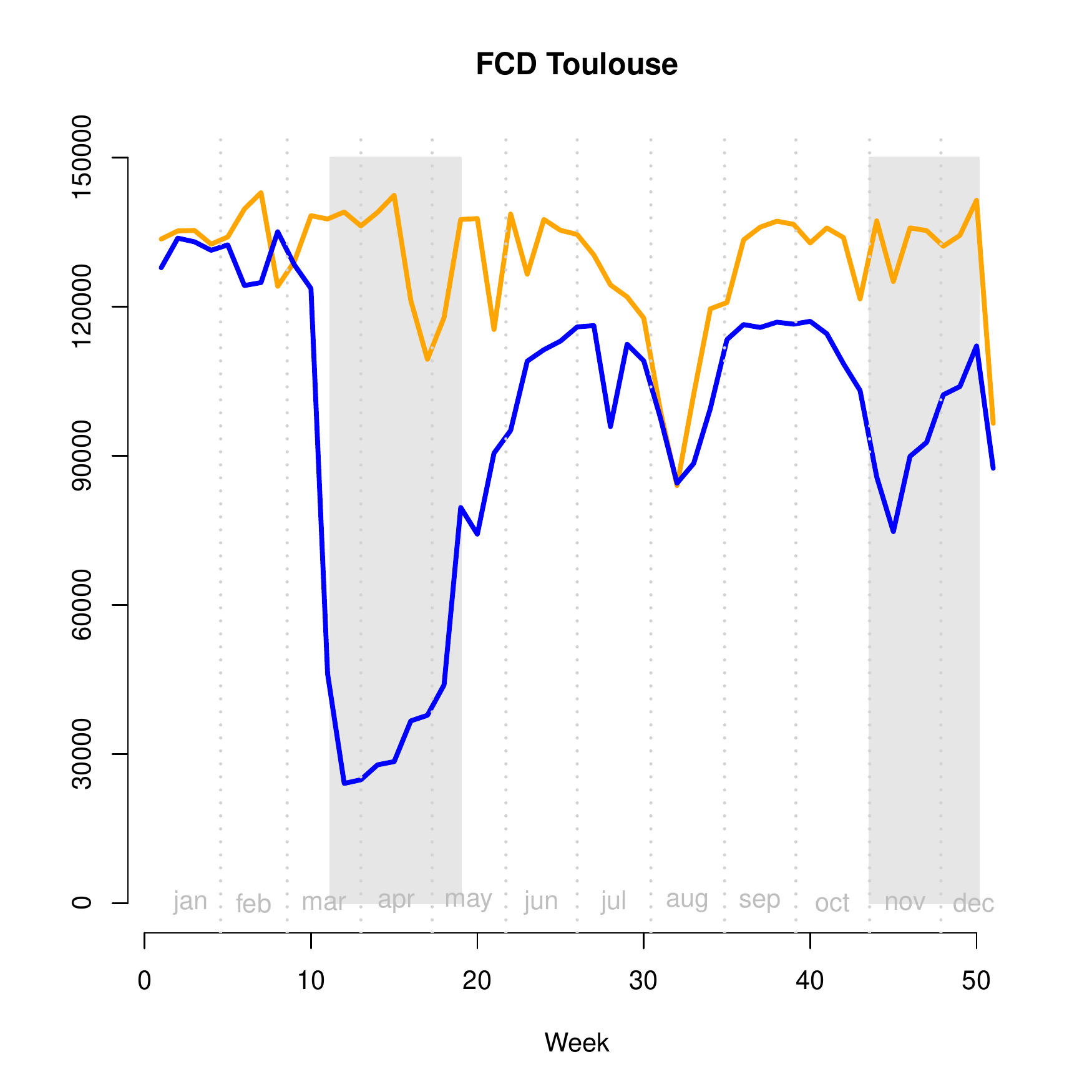}
\includegraphics[width=0.24\linewidth]{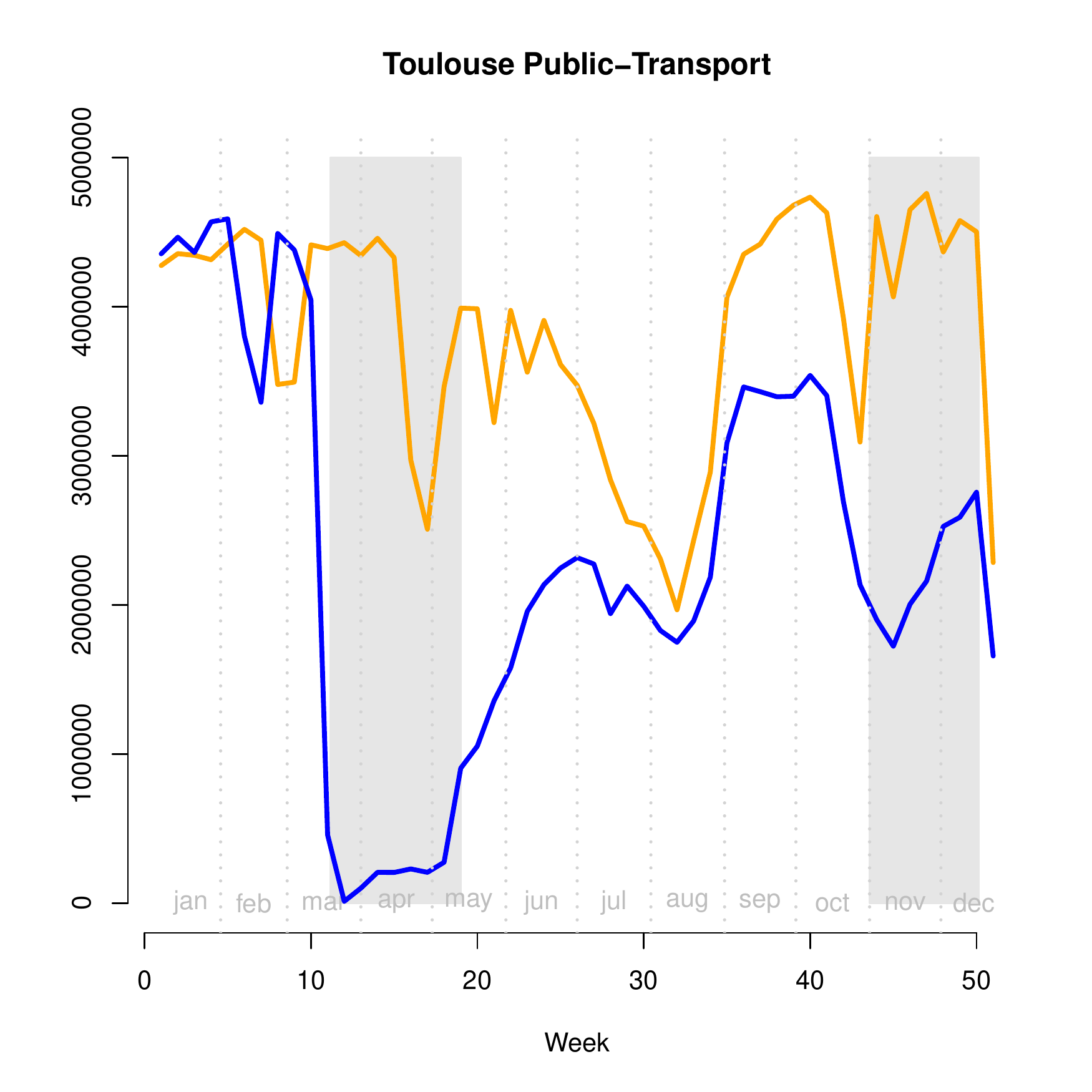}
\caption{(Top) Weekly evolution of BSS use over the course of 2019 and 2020 in Lyon and Toulouse provided by the model. Curves are normalized so that the total sum of use over the year is equal to the observed value for each curve. The lockdown periods in 2020 are represented by gray areas (17 march to 11 may and 30 October to 15 December). (Bottom) Weekly evolution of road traffic (FCD data) and Public-Transport entries, plotted from the raw data. 
}
\label{fig:weekly}
\end{figure}

\Cref{fig:weekly} (top) shows the weekly use of shared bikes for both years and both towns. These two plots are the weekly components of the model. As for the daily component, it differs from the raw data in that weather and holidays effects are corrected, even if they account for little in the explained variance of the data. For a year without pandemic (2019), Toulouse and Lyon curves have the same profile, characterized by a low number of trips during summer holidays followed by a higher number between mid-September and mid-October. The difference between the two extremes varies from nearly one to two in both towns. The end of the year, when many people take vacations, is a second period of low use in both towns. These simple observations show that shared bikes are largely used for working activities, which is reinforced by the magnitude of the morning and evening peaks in Figure \ref{fig:work_day}.  

We have tested a model accounting for different weekly behaviors for working-days and weekend-days, i.e.~with a component $g(\text{week, type of day})$ instead of $g(\text{week})$ in Eq.~\eqref{model_temp} and the additional parameters were not statistically significant. This indicates that any possible weekly evolution difference between working-days and weekend-days is insignificant to explain weekly BSS use along the year. \\

In 2020, rhythms are still similar for both towns but differences exist concerning the pandemic effect. If the number of trips is largely reduced during the lockdowns, the ratio between Lyon and Toulouse remains the same as outside the lockdown with approximately twice as many trips in Lyon. Less strict than the first one (schools remained open and many sectors were able to continue their activity, even if universities switched to online courses), the effect of the $2^{nd}$ lockdown is lower. An interesting period is the one after the first lockdown. Very quickly, in 2 weeks for Lyon and 4 for Toulouse, the number of bike trips exceeds the 2019 levels. This phenomenon is much more important in Lyon where the number of uses even increased by a quarter for some weeks from one year to the other. The jump in BSS usage observed right after the first lockdown, particularly in the city of Lyon, appears to be consistent with previous results for western cities (\cite{bert2020, tokey2020, wang2021}). The difference in the amount of time required to get back to the level of the year 2019 after the $1^{st}$ lockdown in the two towns is due to the decrease of BSS use in 2019 in Lyon in the period between mid-May and mid-June. It corresponds to a special event when, within 2 weeks, a large part of the BSS-bike fleet was stolen. Without this decrease, the amount of time required to get back to the level of 2019 would have been similar in both towns and approximately equal to 1 month. This is similar to what \cite{chai2020} was observed in Beijing but twice shorter than in Brooklyn \cite{wang2021} and in Chicago \cite{hu2021}.  \\

\Cref{fig:weekly} (bottom) allows a comparison with two other types of urban transport, private cars and Public-Transport. Please note that for FCD data, all the urban territory is under consideration, including urban highways around the cities. The first observation is that patterns are very similar in both towns and neither Public-Transport nor road-traffic recovered in 2020 their 2019 levels. The summer period is clearly less impacted in both towns. Public transport's resilience to the pandemic was particularly poor since it remained largely underused until autumn 2020, with patterns for the weeks following the first lockdown which are similar to those during the second lockdown. In both towns, the decrease is about 40\%, as compared to 2019, which is similar to observations in Shenzhen showing a decrease of 34\% \cite{rojas2021}. With a latency of one month as for BSS, road-traffic volumes (as represented by the observed fleet of vehicles in the available FCD data) return to stable levels (outside summer) better than those of Public-Transport and reach around 85\% of the 2019 levels for both towns. The order highlighted here regarding the impact of the pandemic on the three modes of transportation is consistent with previous findings \cite{monterde2022}. Please note that the impact of COVID-19 on these three mobility types in Lyon is very close to what has been observed in Zürich \cite{buchel2022}.

In spite of the high quality of fit, \Cref{fig:weekly} shows high variations of bike usage in 2020 depending on the considered period of the year. So, it is worth looking at how BSS traffic bike is distributed over one typical day for three periods: the first and the second lockdown, and the weeks just following the first lockdown (see \Cref{fig:comparison}). 
The graphs on the left of \Cref{fig:comparison} show that the heavy restrictions of the first lockdown in France almost completely obliterated BSS use. Nevertheless, three of the four peaks can still be perceived and occur at the exact same times, around 8:45 AM and 6:15 PM. 

The weeks following the lockdown reveal different BSS uses in the two towns. In Lyon, the slight morning peak decrease is more than balanced by a late morning and early-afternoon spread-out. It is also worthwhile noticing that, right after lockdown, the evening peak recovered its 2019 level which shows how resilient BSS usage is. In Toulouse, the observed pattern is different, with a volume of usage which does not get back to its pre-pandemic level. The almost-halved morning peak is strongly impacted and this is not balanced by the rest of the day with the three other peaks also largely diminished while inter-peak periods recovered their 2019 level. Under the hypothesis that working from home was equally distributed in Lyon and Toulouse, it suggests that BSS is more resilient in Lyon than in Toulouse for commuting. Comparing these patterns with those of Zurich and Basel \cite{buchel2022} would put Zurich close to Lyon and Basel close to Toulouse. This is consistent with the fact that BSS is slightly more established in Zurich and Lyon than in Basel and Toulouse. 

Finally, the impact of the second lockdown was much less noticeable than the first one with a similar day pattern as usual and only a moderate decrease in volume. Lyon and Toulouse show a very similar pattern. It is worth noticing that, with or without COVID, the morning peak for working-days is, for this period of the year, preceded by a smaller one (around 7:50 AM in Lyon and 8 AM in Toulouse), and this phenomenon is more obvious in Lyon. This small peak, somehow present in the March-May period in 2019, reveals a bimodal distribution which probably deserves being further studied.

\begin{figure}[H]
\centering

\includegraphics[width=0.7\linewidth]{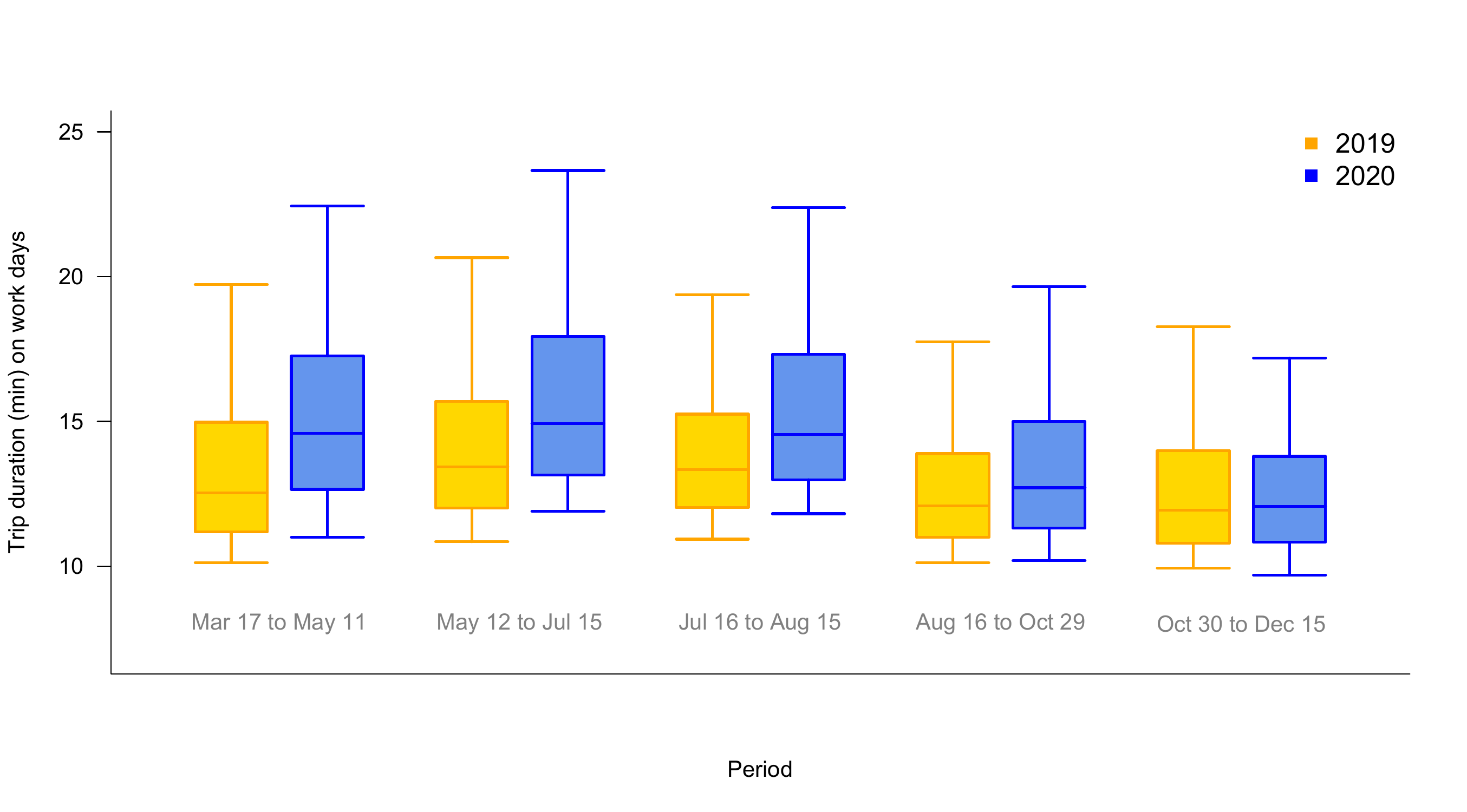}

\vspace{-1.9cm}
\includegraphics[width=0.7\linewidth]{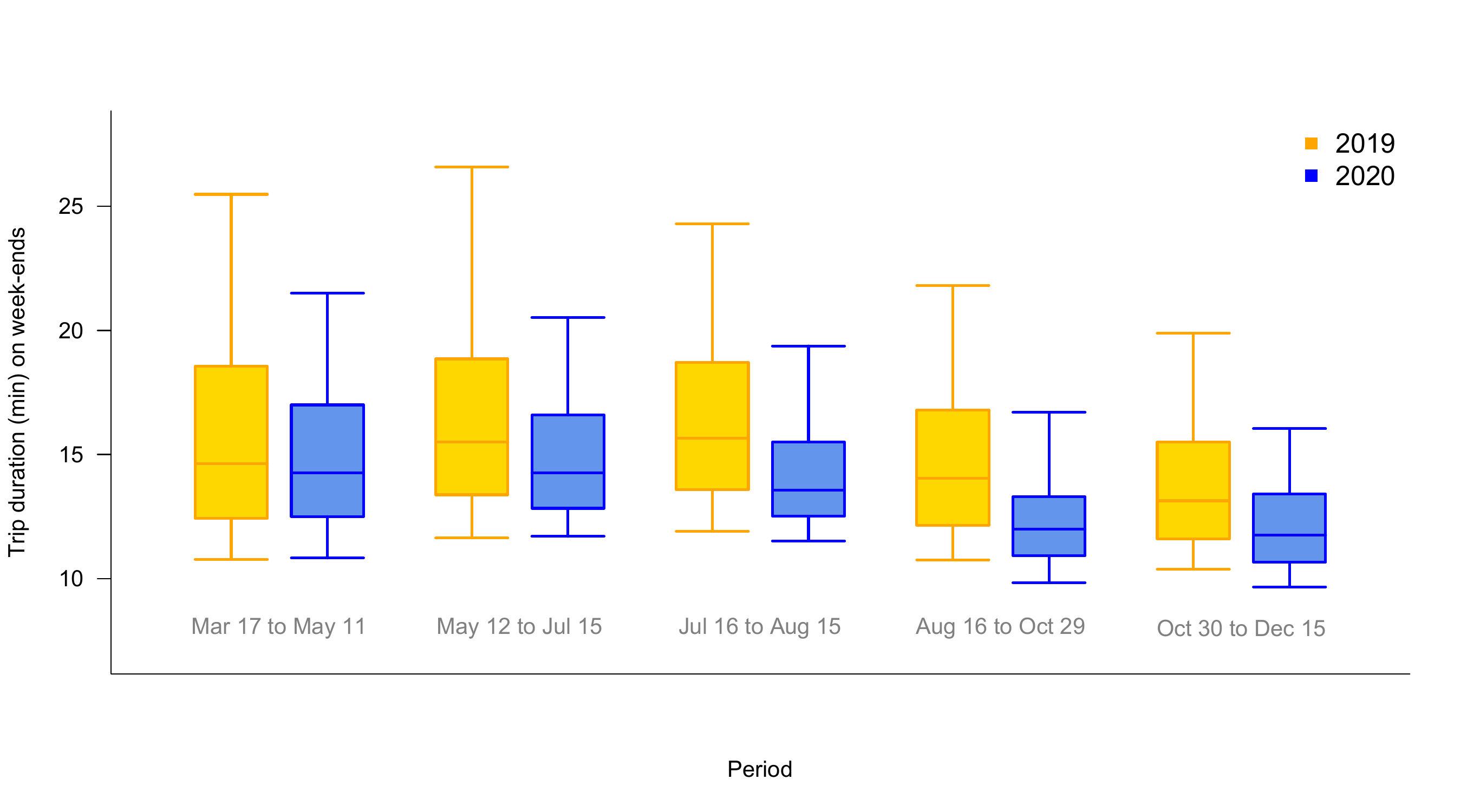}
\caption{Comparison of trips duration over the different time periods for Lyon and Toulouse merged data. First ($Q_1$), second ($Q_2$, median) and third ($Q_3$) quartiles are plotted. The boundaries of the whiskers are the $10^{th}$ and $90^{th}$ percentiles.}
\label{fig:duration}
\end{figure}

Consistent with previous findings \cite{li2020, teixeira2020, hu2021, rojas2021}, trip duration increased during the pandemic situation, as pictured in \Cref{fig:duration}. However, this overall increase masks opposite situations between working and weekend days, which have been little documented so far. 
Distributions are highly skewed due to the presence of a few long (more than $2$ hours) trips, not visible in the boxplots. An analysis of variance has been conducted on trip duration trimmed of values below one minute and above two hours in order to handle the outliers and skewness. It shows that the differences in trip duration between 2019 and 2020 are highly significant for working-days on the first four periods but becomes less visible in the second lockdown (p-value $\approx 0.011$).

Weekends display the opposite. While distribution shifts toward longer trips for working-days, it shifts toward shorter trips for weekend-days. Long weekend trips tend to disappear in 2020, as opposed to longer commutes. These observations constitute a major difference when compared to Zurich for which \cite{li2020} reported an increase in duration also during weekends with a completely different behavior. However, our results are in line with \cite{teixeira2020} data for New-York, readable in some of their figures and tables.

\section{Main spatio-temporal patterns of hourly BSS use}
\label{sec:efa}

In this section, we used Exploratory Factor Analysis (EFA) \cite{Jolliffe2002} to investigate the spatio-temporal effects of the pandemic on BSS usage. 
EFA is an unsupervised data compression technique that allows extracting from data a small number of latent factors that meaningfully summarize the covariance matrix of the observed features (i.e., variables). 

We chose to extract two distinct EFA models from the available datasets, one for Toulouse and one for Lyon, each covering the two years 2019 and 2020. In both models, the observed variables correspond to the $p=287$ (resp. $p=430$) dock-stations in Toulouse (resp. Lyon), for which we observe the number of trips which begin at a given station in a given one-hour slot. Therefore, for each station, we have a total of $q=17544=24 \times (365+366)$ observations. Each one-hour time slot represents a population sample (i.e., an individual) on which all variables are observed.
Unlike our previous analysis in which we distinguished between working-days, weekend-days or periods of the year on the basis of prior knowledge of relevant factors, we now assume that we do not have such preliminary knowledge, and thus aim to learn directly from the data the factors that can best explain existing similar temporal patterns between the different stations. 

In the fundamental equation of the EFA model, observed variables are expressed as linear combinations of a given (fixed) number of (unknown) common factors plus an additive term that can be considered as noise, or, in EFA terminology, as the uniqueness associated to each observed variable. Based on such principles, EFA fundamentally differs from Principal Component Analysis \cite{dewinter2016} in the sense that factors are constructed by assuming that the covariance of the observed variables could be decomposed into two parts, i.e., a common one and a unique one, while in PCA all variance in the observed variables is analyzed and used for the determination of the components. In other words, PCA aims to find components that allows reproducing (exactly) the observed variance as linear combinations of the identified components, while in EFA focuses on the interpretation of the relationships among the variables and on identifying the latent factors which explain such relationships as noisily observed in the data covariance matrix.

Following the usual EFA terminology, we call \textit{(common) factors} the set of K latent ``variables'' which makes up the construction of the observed variables and \textit{factor loading matrix} the set of numerical relationships describing how much each factor explains each (observed) variable. The fundamental equation of factor analysis is thus the following:
$$\mathbf{X=\Lambda F+U}$$ where $\Xm$ is the $p$-vector of observed variables, $\Fm$ is the $K$-vector of unknown factors,  $\Lambdam$ is the unknown $p \times K$ matrix of factor loadings and $\Um$ is the unknown $p$-vector of error variables (uniqueness) associated to each of the $p$ observed variable. 

Factor loadings can be interpreted as correlations between factors and observed variables (the stations in our case) and range from $-1$ to $1$. Loadings close to $-1$ or $1$ indicate that the factor strongly influences (negatively or positively, respectively) the observations (time dynamics in our case) of the station. Loadings close to $0$ indicate that the factor has little influence on the station. For each loading and for both towns, we can draw a map where each dock-station is geographically located with a circle whose color is function of the loading value. These maps will be called ``loading maps''.

Each entry of $\Um$ is called a \textit{unique factor}. Since each unique factor is specific to one variable, the error terms are independent and $\Cov(\Um)$ is a diagonal matrix. 

As part of the EFA methodology, it is possible to estimate (e.g., via regression analysis) \textit{factor scores}, i.e., the $q$-sized vector of coefficients for each of the K factors over the sampled population. Factor scores can have negative and positive values and are usually standardized. In our case study, the scores for a given factor represent the evolution of that factor over the one-hour time slots related to the two years 2019 and 2020, and thus allow visualizing the temporal dynamics of BSS usage for the set of dock stations highly represented by the considered factor (i.e., stations exhibiting high loadings on that factor).

More details on the EFA model, the solving procedure to compute factor loadings and uniqueness, as well as approaches to select the number of $K$ factors and to estimate factor scores can be found in the well-established literature on the subject \cite{Mulaik2009,comrey2013,joreskog1978}. 

The following analysis has been performed using the departure data, but a similar one on the arrivals has been conducted, producing identical conclusions. Therefore, only the data from departure are presented to avoid redundancy. As previously discussed in relation to factor scores, time series have been centered and scaled during the factor analysis process, which means that the comparisons between stations is based on their pace and not their intensity.

\subsection{Technical choices}
Factor extraction was performed using the 
Minimum Residuals (MINRES) approach \cite{harman1966}. The working principle of MINRES is to minimize the sum of off-diagonal squared residuals of the correlation matrices of $\Xm$ (the observed one and that reproduced from the identified common factors). 
Unlike maximum likelihood estimation \cite{lawley1940}, MINRES does not rely on any assumption about the distribution of observed variables and can produce valid solutions even when applied to singular matrices~\cite{joreskog1978, briggs2003}.

An important design choice concerns the number of common factors that EFA should target. We rely on Parallel Analysis (PA) \cite{horn1965}, which uses the eigenvalues of the observations' correlation matrix as rough estimates of the actual common factors. Specifically, PA compares such eigenvalues with those obtained from a random correlation matrix for which no factors are assumed. 
The presence of common factors shall induce large eigenvalues: the number of factors is set to the lowest rank above which all data eigenvalues are larger than those from the uncorrelated variables. In our case study PA yielded 23 (resp. 27) factors for Toulouse (resp. Lyon). They account for 41.5\% (resp. 53.6\%) of the total variance.

Once a set of factors is extracted, it is usual to perform a rotation in order  to produce a more interpretable and simplified solution \cite{abdi2003}. Varimax rotation over BSS data worked best in that respect. It is an orthogonal rotation (which thus keeps factors uncorrelated) which maximizes the sum of the variance of the squared loadings.

In the following, factors are ordered in decreasing order of the variance they explain. Regarding the variance explained by the 22 and 26 common factors respectively, the first 3 factors for Toulouse and first 4 factors for Lyon account for 82\% of this variance in each case. For both Toulouse and Lyon, two more factors are of interest  by their particular spatio-temporal distribution. Lower-ranked factors are either restricted to a handful of stations or concentrated in a very small time period (e.g., a particular day). 
Therefore, in the following we will consider only 5 factors for Toulouse and 6 for Lyon. 
The $3^{rd}$ factor in Lyon and the $4^{th}$ factor in Toulouse, which are representative of the new bike stations installed by end of 2019 / beginning of 2020, will not be discussed here as they are deemed to be irrelevant to the scope of the paper.

\subsection{Factors for Lyon vs Factors for Toulouse}
\label{subsec:NewResults}

In the loading maps, only stations with loadings greater than .2 are plotted.

\subsubsection{The morning peak}

%%%%%%%%
\begin{figure}[H]
\centering
\sbox{\tempbox}{
\includegraphics[width=0.32\textwidth]{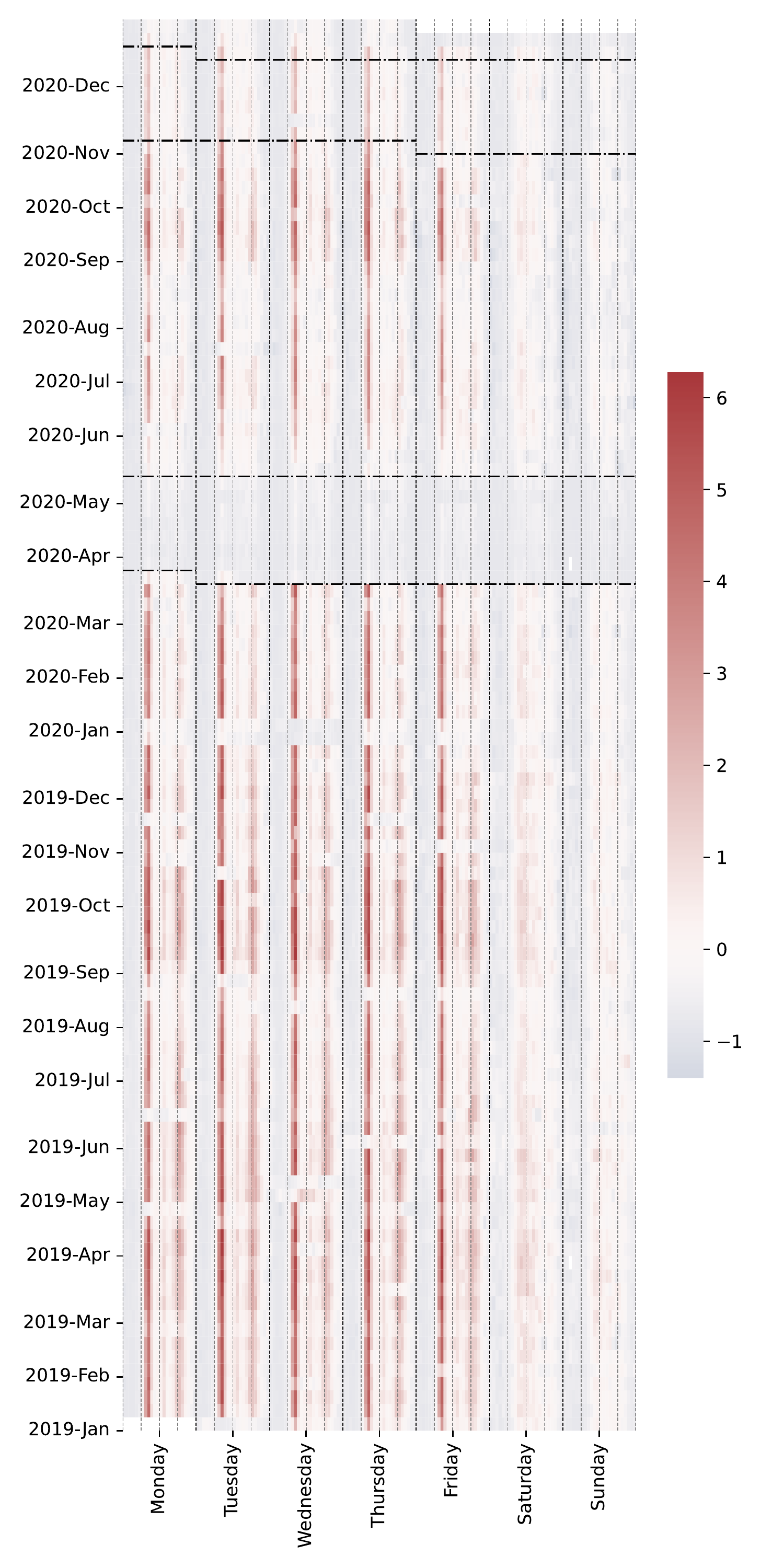}
\includegraphics[width=0.32\textwidth]{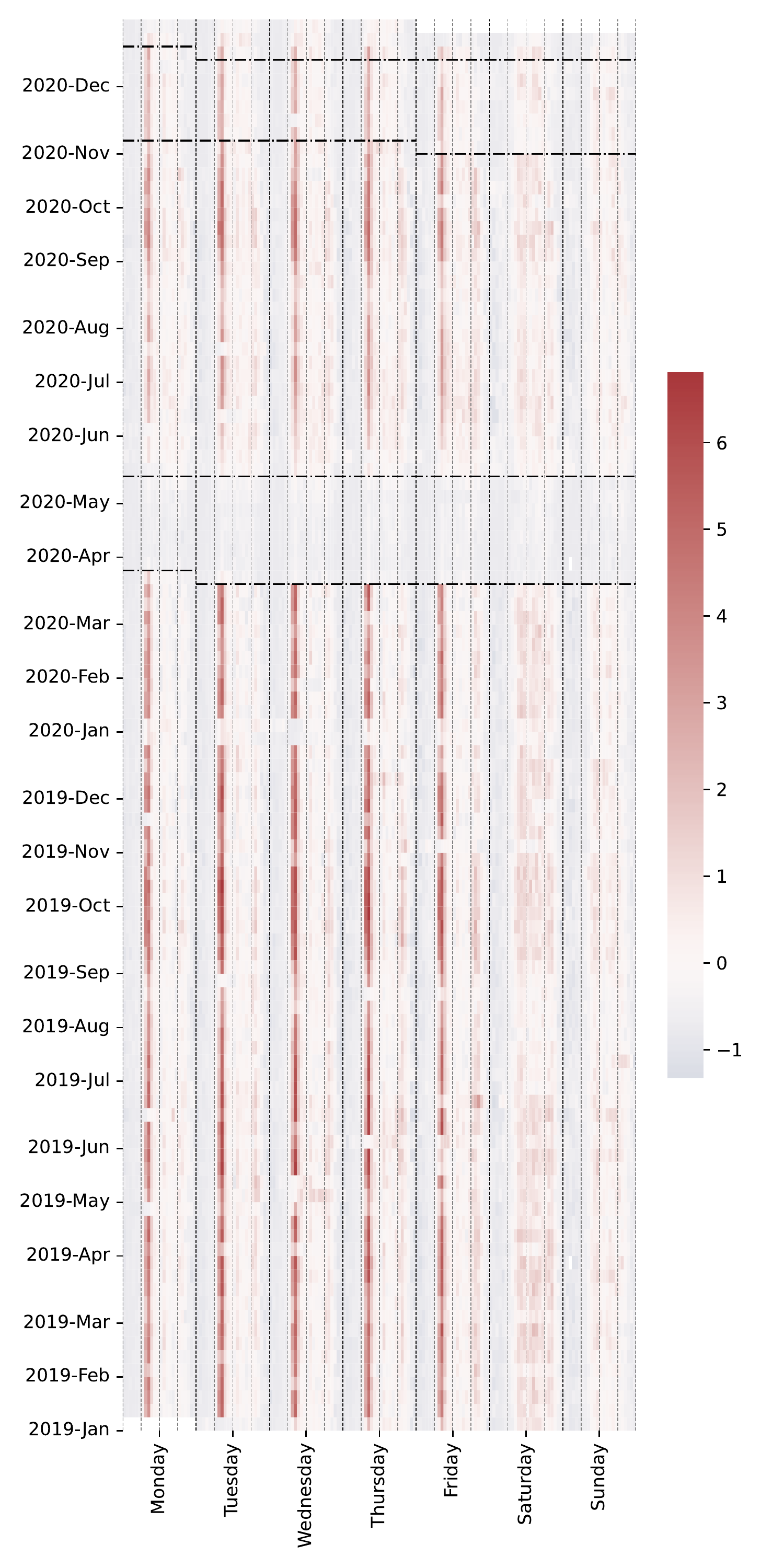}
}
\subcaptionbox{Hourly representation of the $1^{st}$ factor scores - Lyon (left) and Toulouse (right)}{\usebox{\tempbox}}
\parbox[b][\ht\tempbox][s]{0.3\textwidth}{%
  \subcaptionbox{$1^{st}$ factor Loading Map - Lyon}
{\includegraphics[width=.26\textwidth]{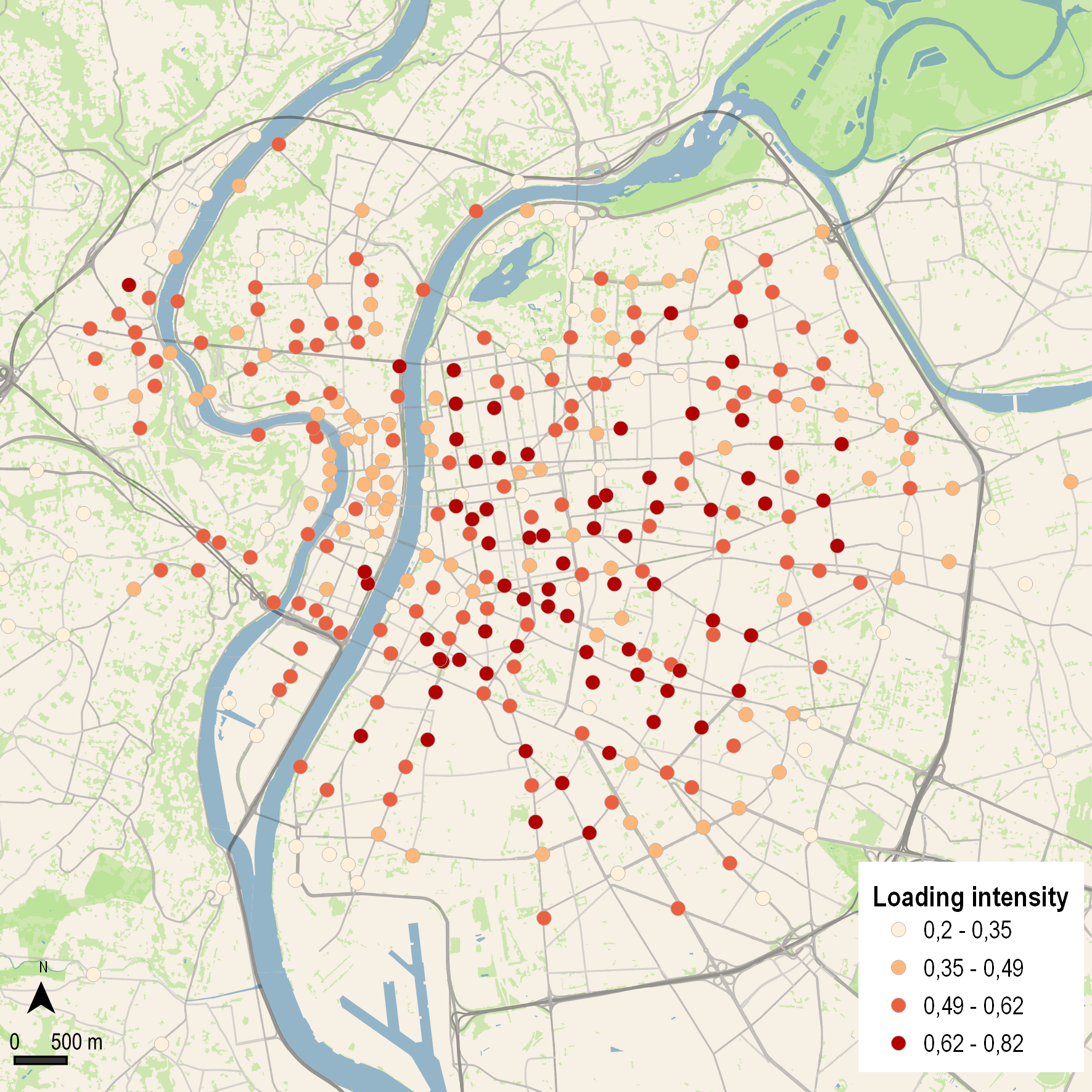}}

\vspace{.1cm}

\subcaptionbox{$1^{st}$ factor Loading Map - Toulouse}
{\includegraphics[width=.26\textwidth]{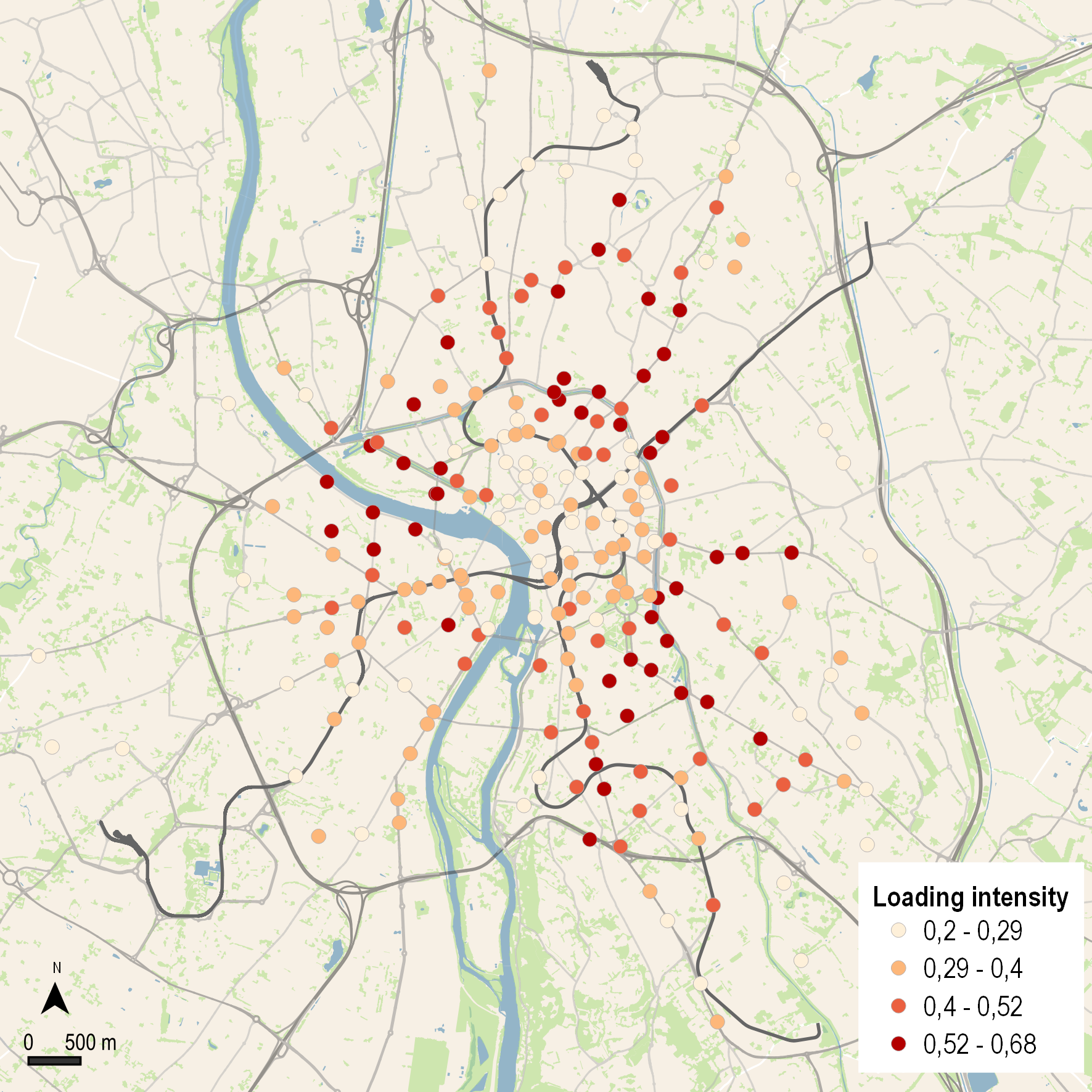}}
}
\caption{The EFA factors representing the morning peaks in Lyon and Toulouse.}
\end{figure}
%%%%%%%%%
The first factor points out a dominant morning peak for both Lyon and Toulouse for working-days in 2019 and 2020, which completely disappears during the first lockdown. For both towns, it is the only factor that shows significant scores for this morning peak and for a large number of stations. 
This factor is mainly associated with departures from residential areas. This phenomenon is particularly observable in the city of Toulouse where residential areas are more clearly separated from business districts and activity centers. Loadings for the city of Lyon appear to be lower in the most touristic zones, such as the ``Presqu'Ile'', and in La Doua university district. This can be explained by considering that such locations have very particular traffic patterns at this time of the day.   
It is worth noticing that, in the city of Lyon, the first factor also shows low but significant positive scores for the end-of-afternoon peak, which is not the case for Toulouse. This can be again explained in terms of the urban fabric associated to the corresponding areas of the city of Lyon, which are characterized by a more mixed residential-work land use than in Toulouse. As a consequence, dock stations in these zones show some non-negligible activity (bike trip departures) also at the end of the afternoon, probably due to people leaving their work places at the end of the day.

\subsubsection{The end-of-afternoon peak}

%%%%%%%%%%%%
\begin{figure}[H]
\centering
\sbox{\tempbox}{
\includegraphics[width=0.32\textwidth]{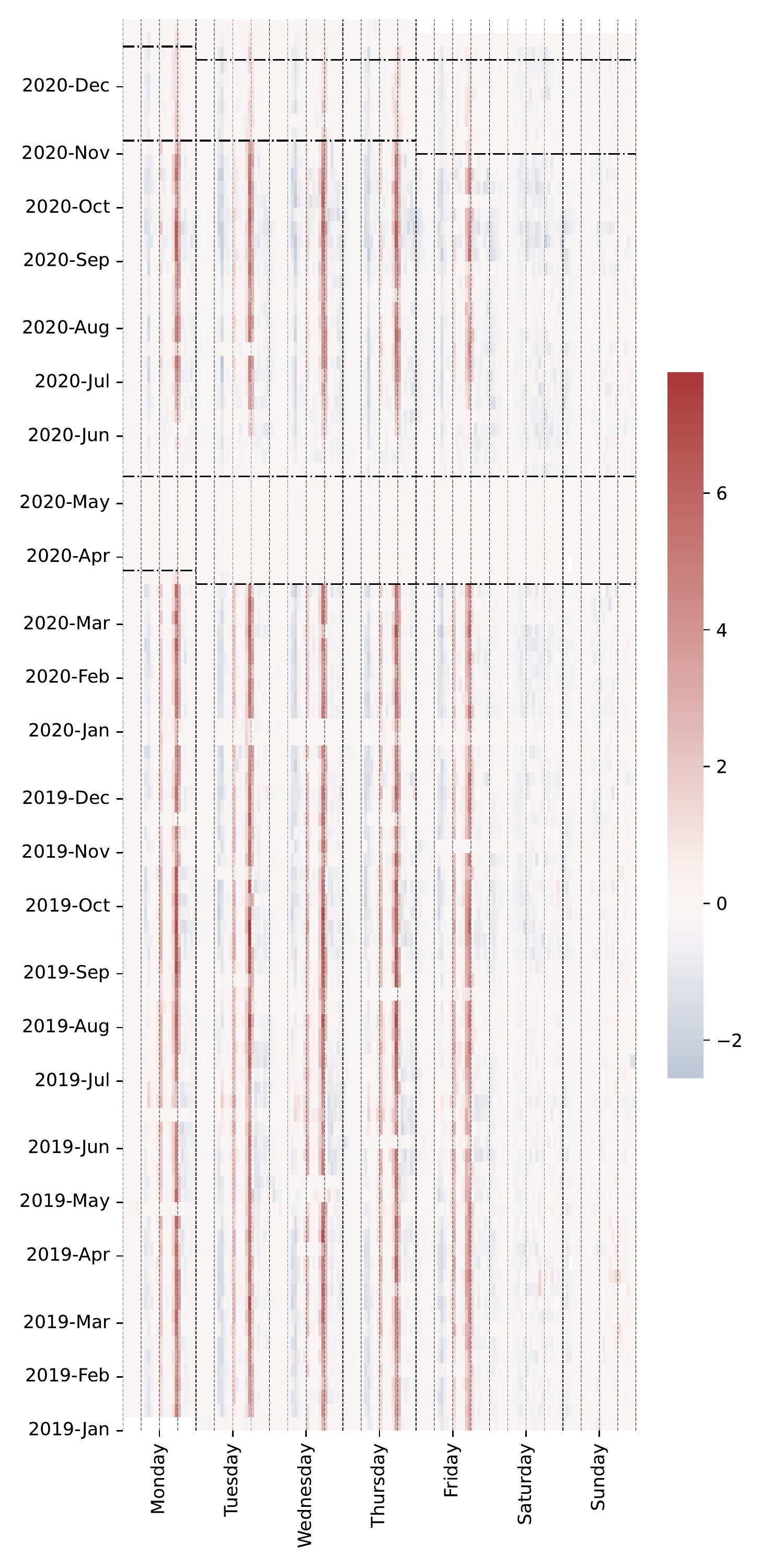}
\includegraphics[width=0.32\textwidth]{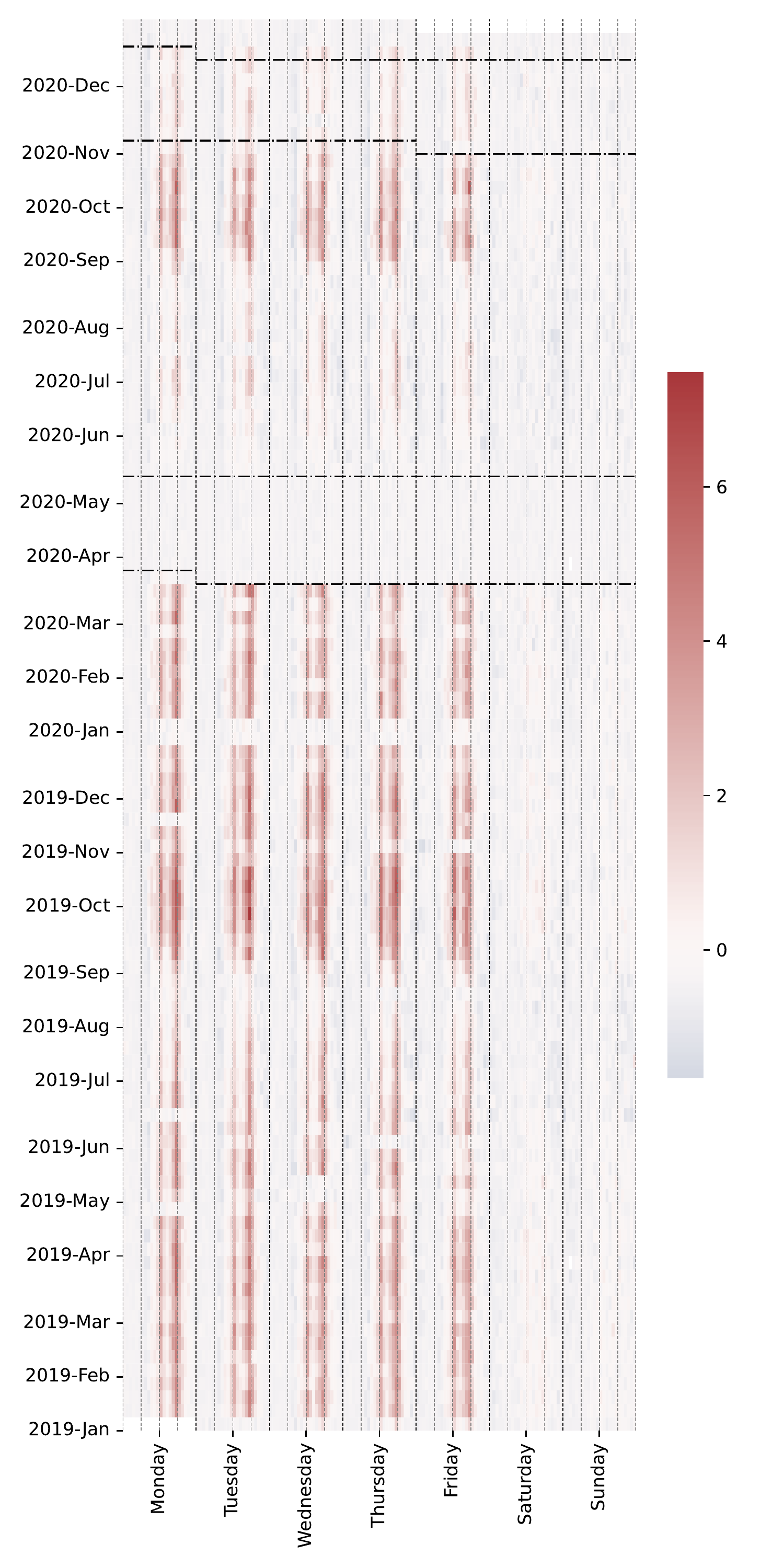}
}
\subcaptionbox{Hourly representation of the $4^{th}$ factor scores for Lyon (left) and $2^{nd}$ factor scores for Toulouse (right)}{\usebox{\tempbox}}
\parbox[b][\ht\tempbox][s]{0.3\textwidth}{%
  \subcaptionbox{$4^{th}$ factor Loading Map - Lyon}
{\includegraphics[width=.26\textwidth]{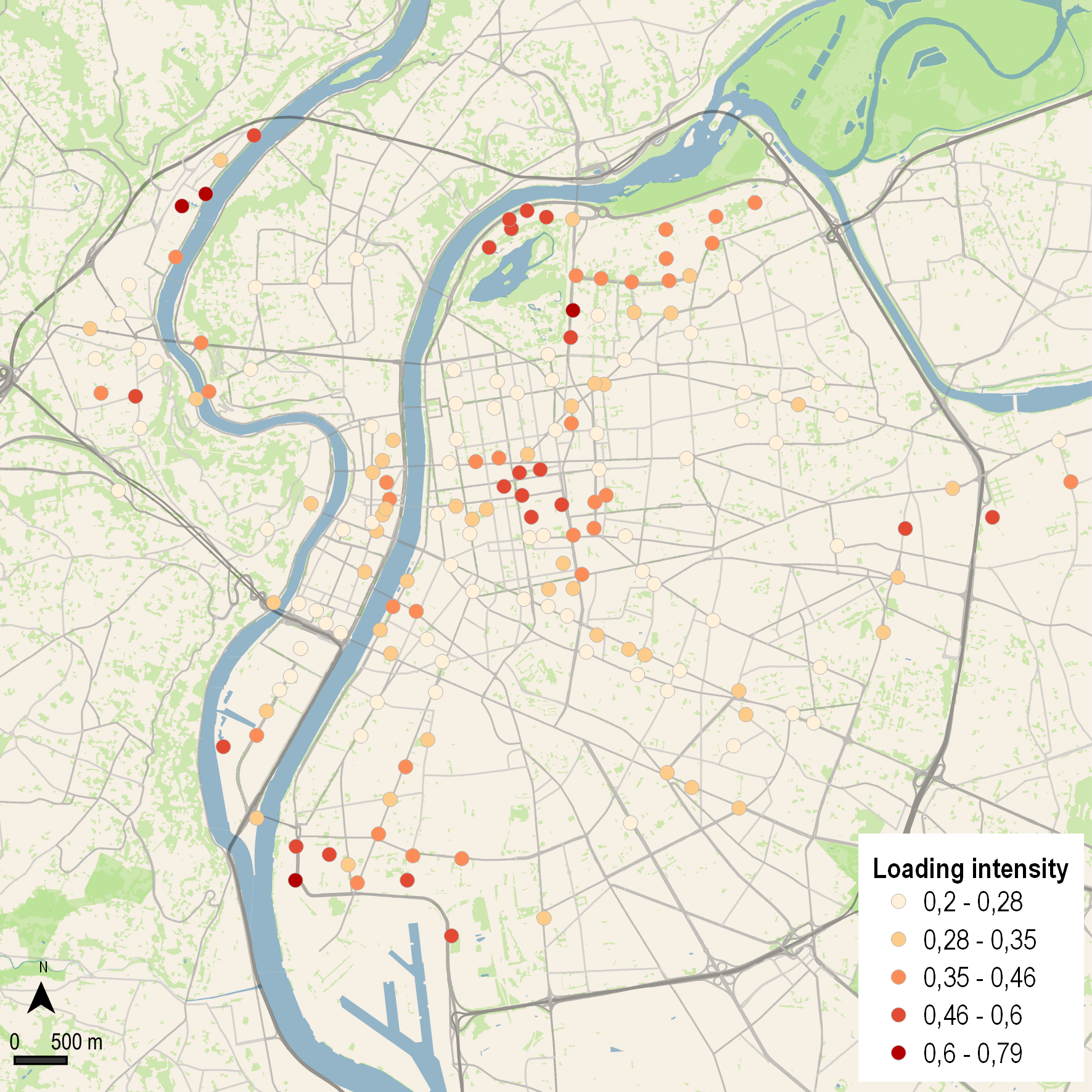}}

\vspace{.1cm}

\subcaptionbox{$2^{nd}$ factor Loading Map - Toulouse}
{\includegraphics[width=.26\textwidth]{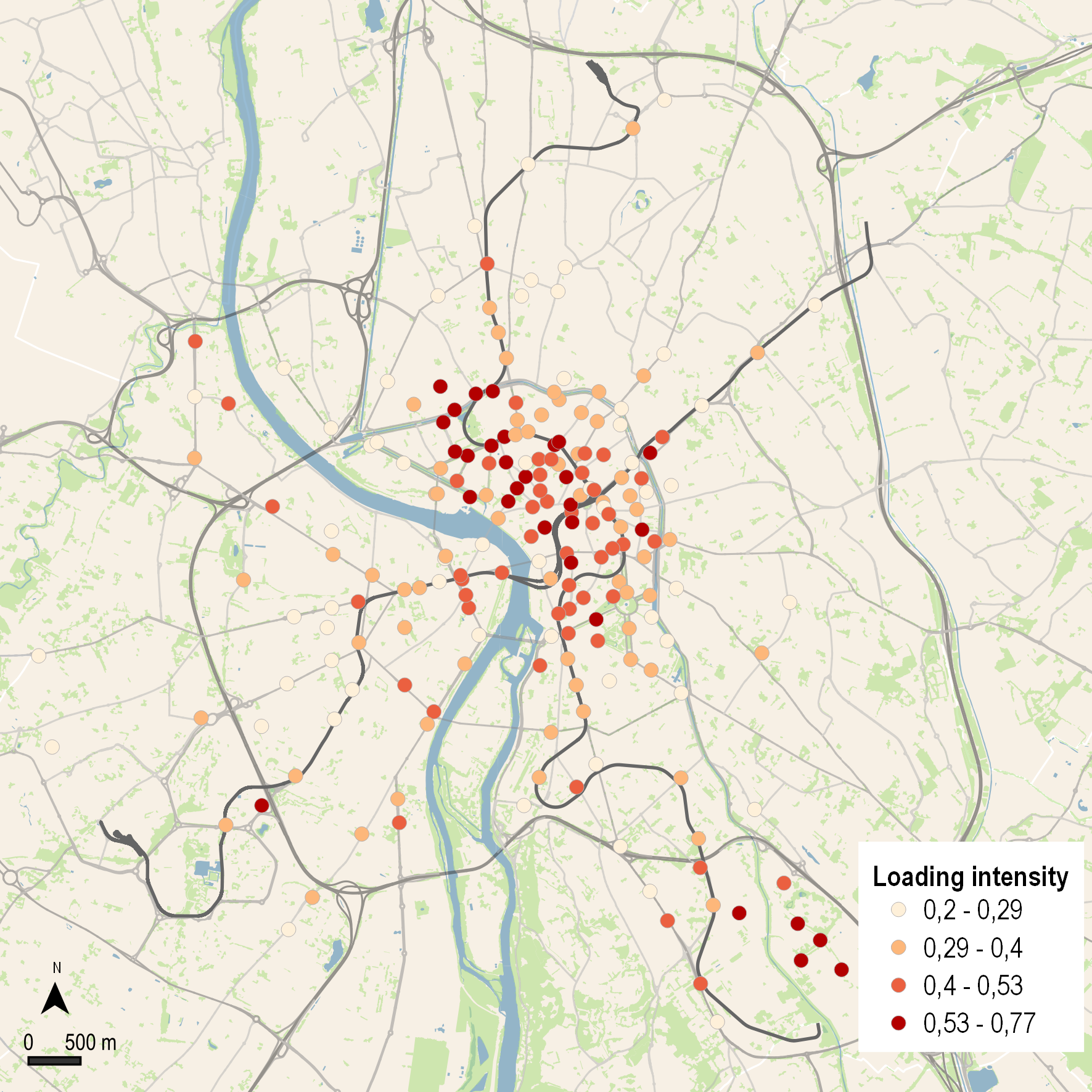}}
}

\caption{The EFA factors representing the end-of-afternoon peak in Lyon and Toulouse.}
\end{figure}
%%%%%%%%%%%%%%
Factor 4 for Lyon and factor 2 for Toulouse appear to be representative of the end-of-afternoon peak particularly related to work places during the 5 working-days. Several differences in this end-of-the-afternoon peak can be highlighted between the two towns. First of all, as one of the effects of the COVID-19 pandemic, the end-of-afternoon traffic peak in 2020 took longer to get back to pre-pandemic levels in Toulouse than in Lyon. This is probably partly due to the important role that university student-related usage play in this factor for Toulouse (the loading map show that dock-stations near the universities are particularly active) and the fact that a lot of them left the city of Toulouse to go back to their home cities and benefit from online courses. Secondly, we can notice that this end-of-afternoon factor is more concentrated, both in time and space, for the city of Lyon. Stations concerned by factor 4 for Lyon are in fact located in the main work places of the city, while the whole city-center (plus the university campuses) is concerned by factor 2 in Toulouse. This can be explained again in terms of the different fabric of the two cities and the fact that the end-of-afternoon peak related to business activities is already partially represented by factor one, together with the morning peak, for stations belonging to areas with a more mixed residential-work behaviors.

\begin{figure}[H]
     \centering
     \begin{subfigure}[b]{0.4\textwidth}
         \centering
         \includegraphics[width=\textwidth]{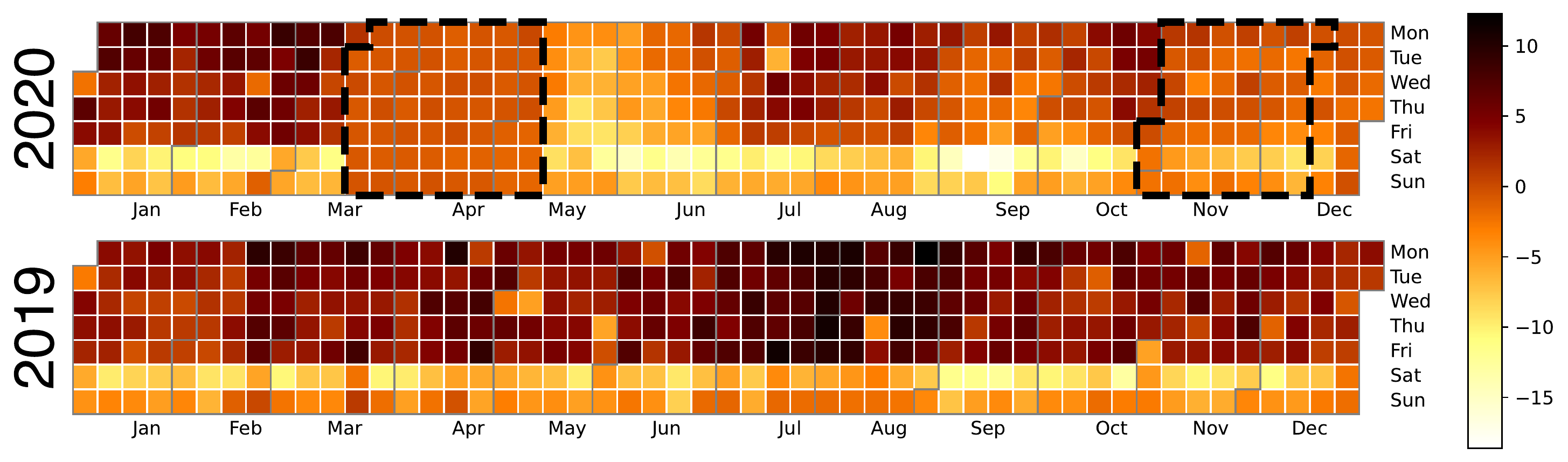}
         \caption{Lyon ($4^{th}$ factor)}
     \end{subfigure}
     %\hfill
     \begin{subfigure}[b]{0.4\textwidth}
         \centering
         \includegraphics[width=\textwidth]{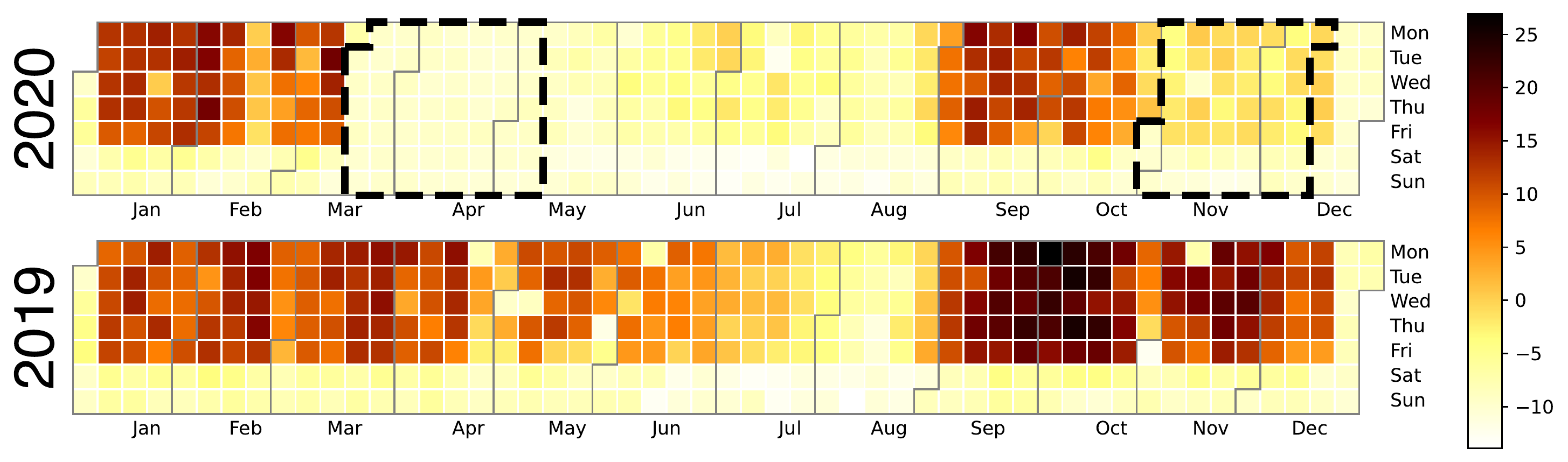}
         \caption{Toulouse ($2^{nd}$ factor)}
     \end{subfigure}
\caption{Daily sum of all station scores over the $4^{th}$ factor for Lyon and $2^{nd}$ factor for Toulouse.  }
\label{fig:working-weekend}
\end{figure}

\begin{figure}[H]
\centering
\sbox{\tempbox}{
  \includegraphics[width=0.25\textwidth]{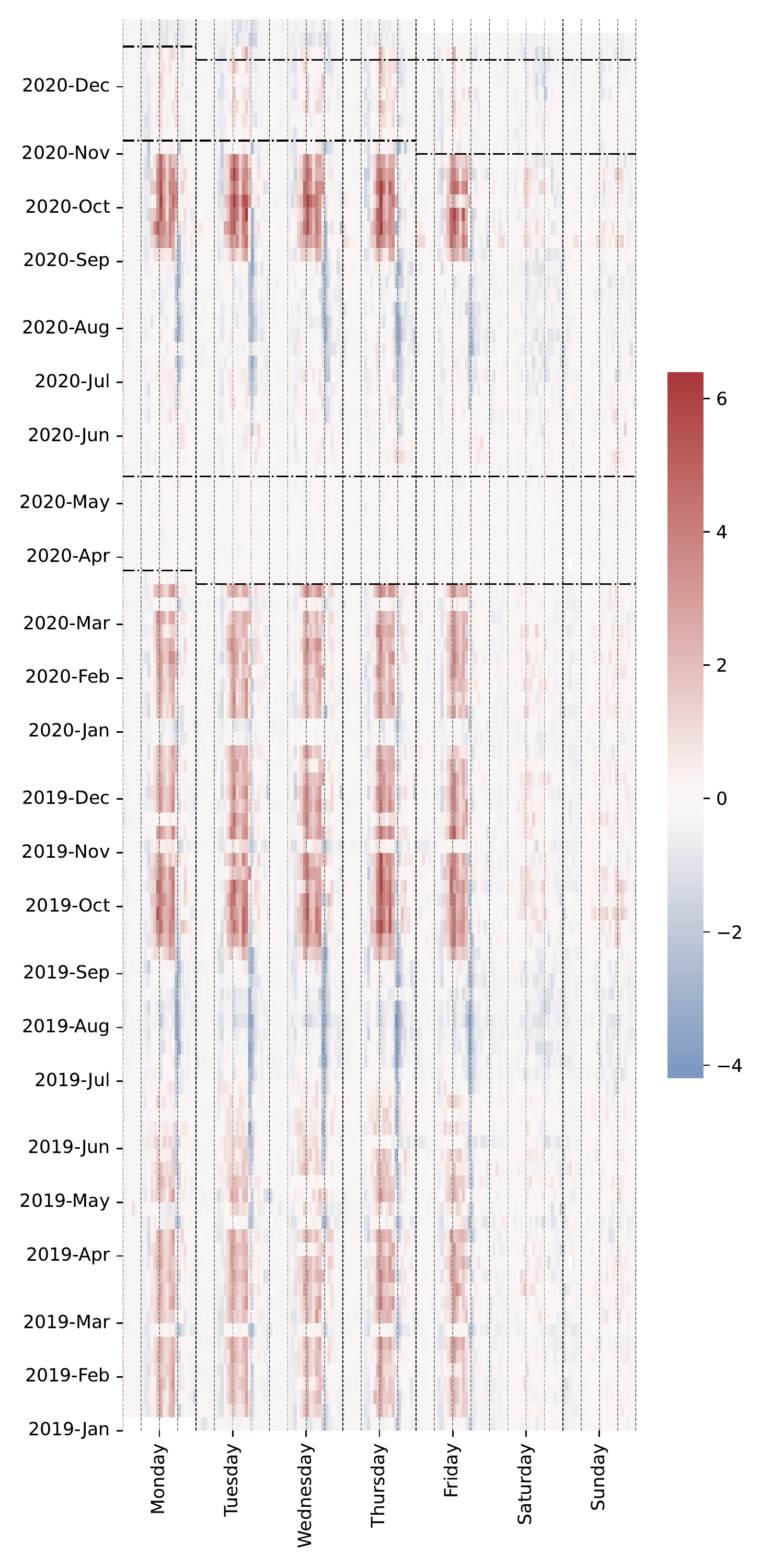}%
}
\subcaptionbox{Hourly representation of the factor scores}{\usebox{\tempbox}}~
\parbox[b][\ht\tempbox][s]{0.55\textwidth}{%
\centering
  \subcaptionbox{Daily representation of the factor scores}{\includegraphics[width=.45\textwidth]{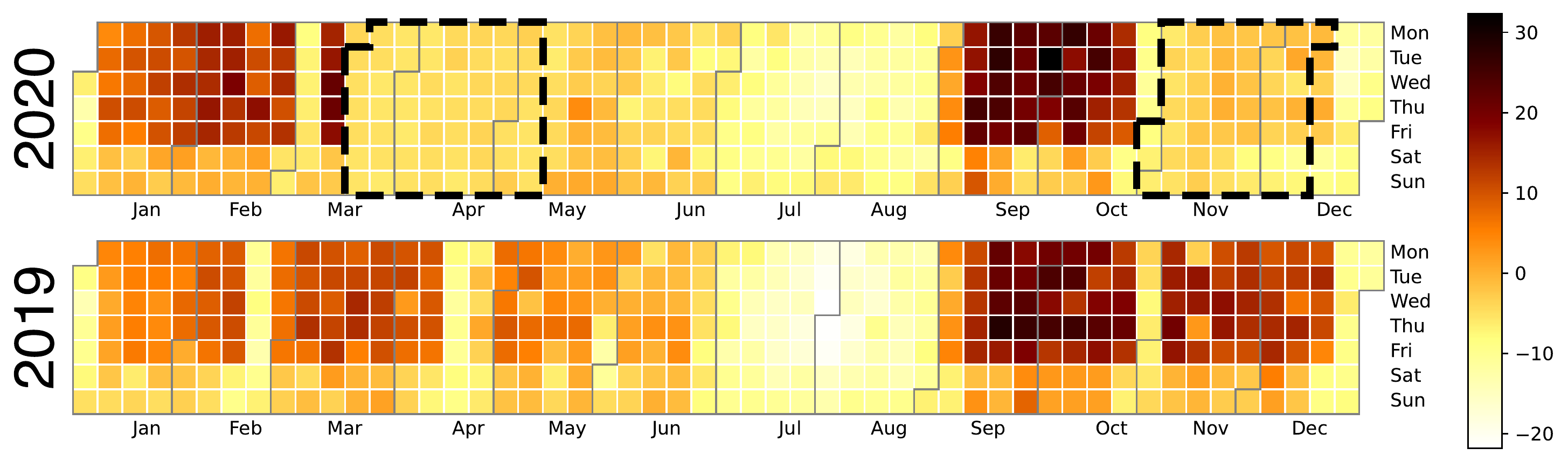}}
  \subcaptionbox{Loading Map}{\includegraphics[width=.3\textwidth]{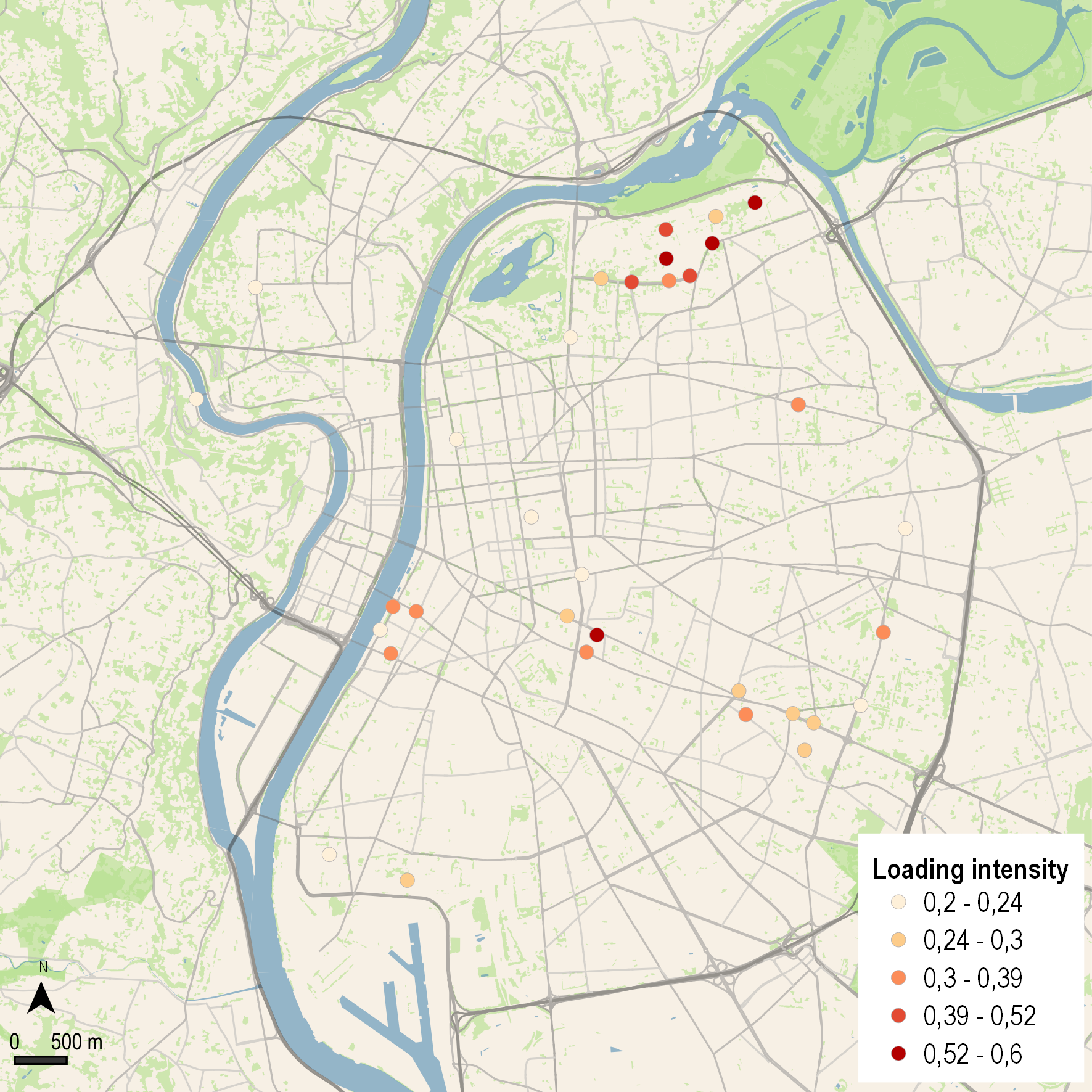}}
}\qquad
\caption{The $5^{th}$ EFA factor representing university student BSS usage in Lyon.}
\label{fig:5_factor_Lyon}
\end{figure}

Finally, among all the factors, these factors are the only ones that count positively for working-days (days with a positive score sum) and negatively for weekend-days (days with a negative score sum). \Cref{fig:working-weekend} shows that these two factors have pretty different time profiles. It is difficult to explain such differences in more detail without further study, but comparisons with university holidays appear to confirm the fact that student usage plays a fundamental role in this Toulouse BSS factor profile. In this respect, the performed EFA analysis does not bring out a specific factor for student profile of BSS usage in Toulouse, but it detects one such factor for Lyon (\Cref{fig:5_factor_Lyon}). The loading map corresponding to factor 5 for Lyon is in fact perfectly associated to university campuses, and university holidays or online courses restriction period due to the COVID pandemic.   

\subsubsection{The late-evening and night-time factor}

%%%%%%%%%%%
\begin{figure}[H]
\centering
\begin{subfigure}[b]{0.55\textwidth}
    \centering
    \includegraphics[width=.49\textwidth]{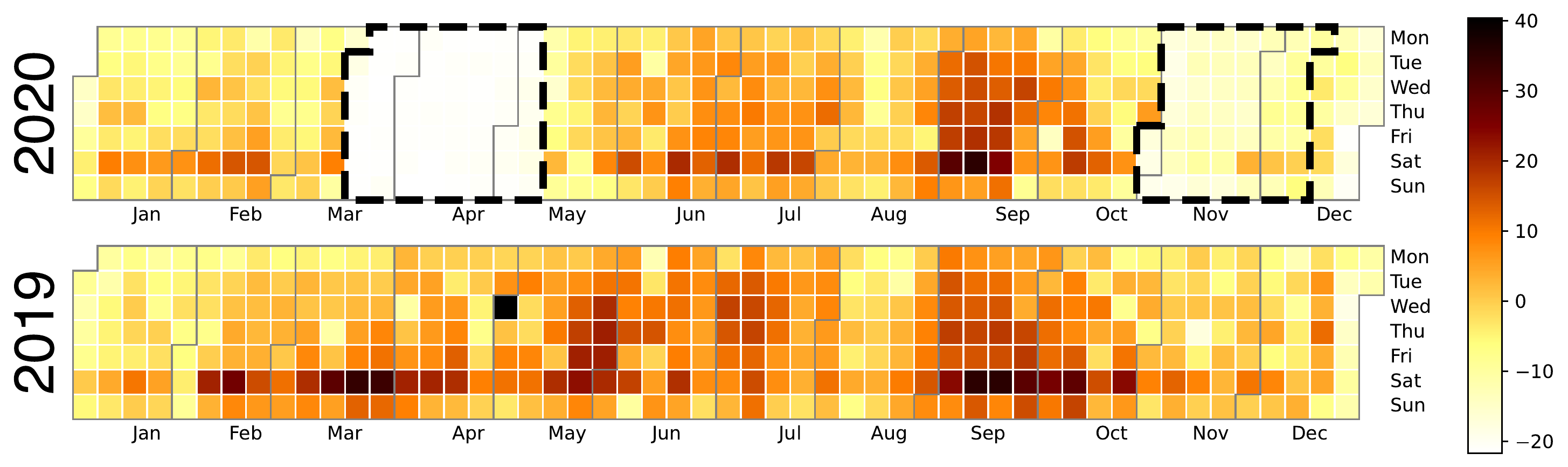}
    \includegraphics[width=.49\textwidth]{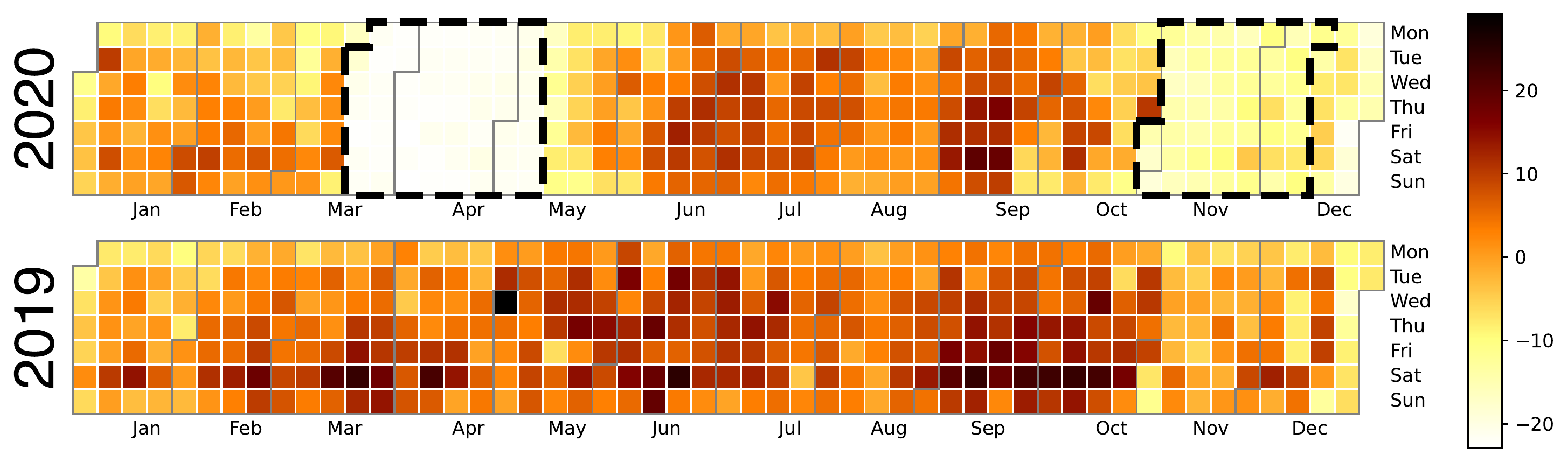}
    \includegraphics[width=.49\textwidth]{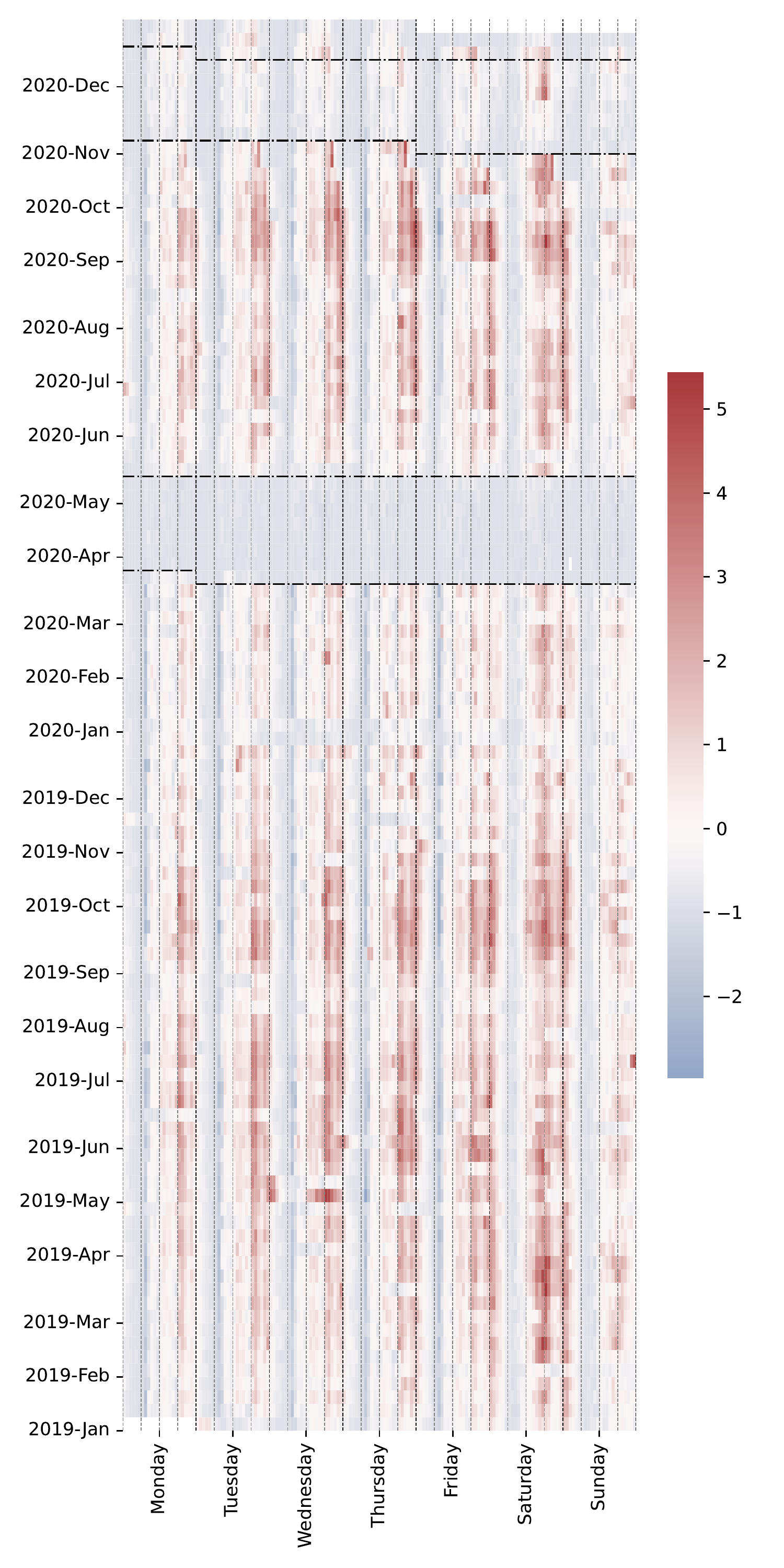}
    \includegraphics[width=.49\textwidth]{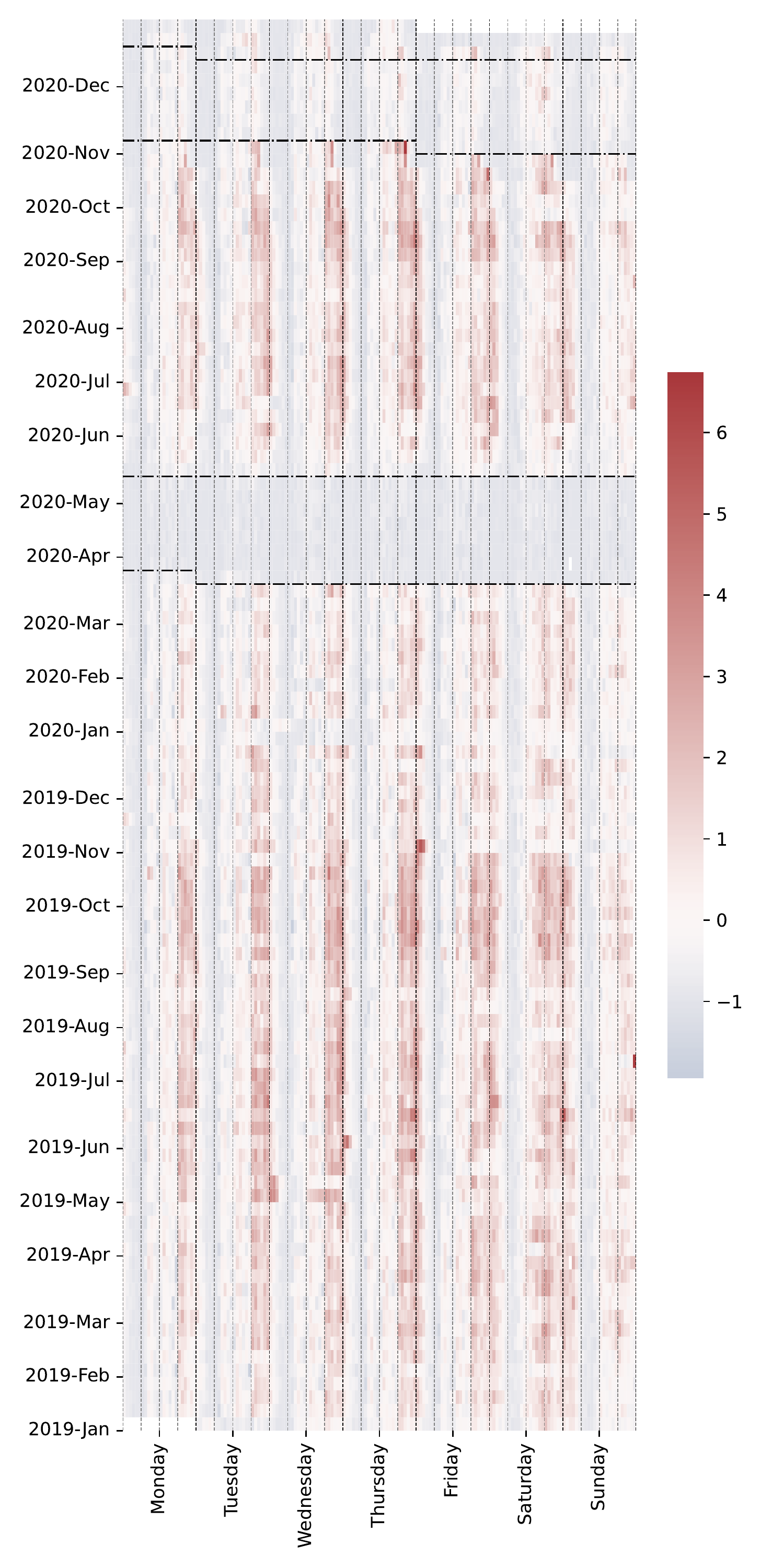}
    \subcaption{Daily and hourly representation of the $2^{nd}$ factor scores for Lyon (left) and $3^{rd}$ factor scores for Toulouse (right)}
    \label{fig:my_label}
\end{subfigure}
\hspace{1cm}
\begin{subfigure}[b]{0.35\textwidth}
    \centering
    \includegraphics[width=.8\textwidth]{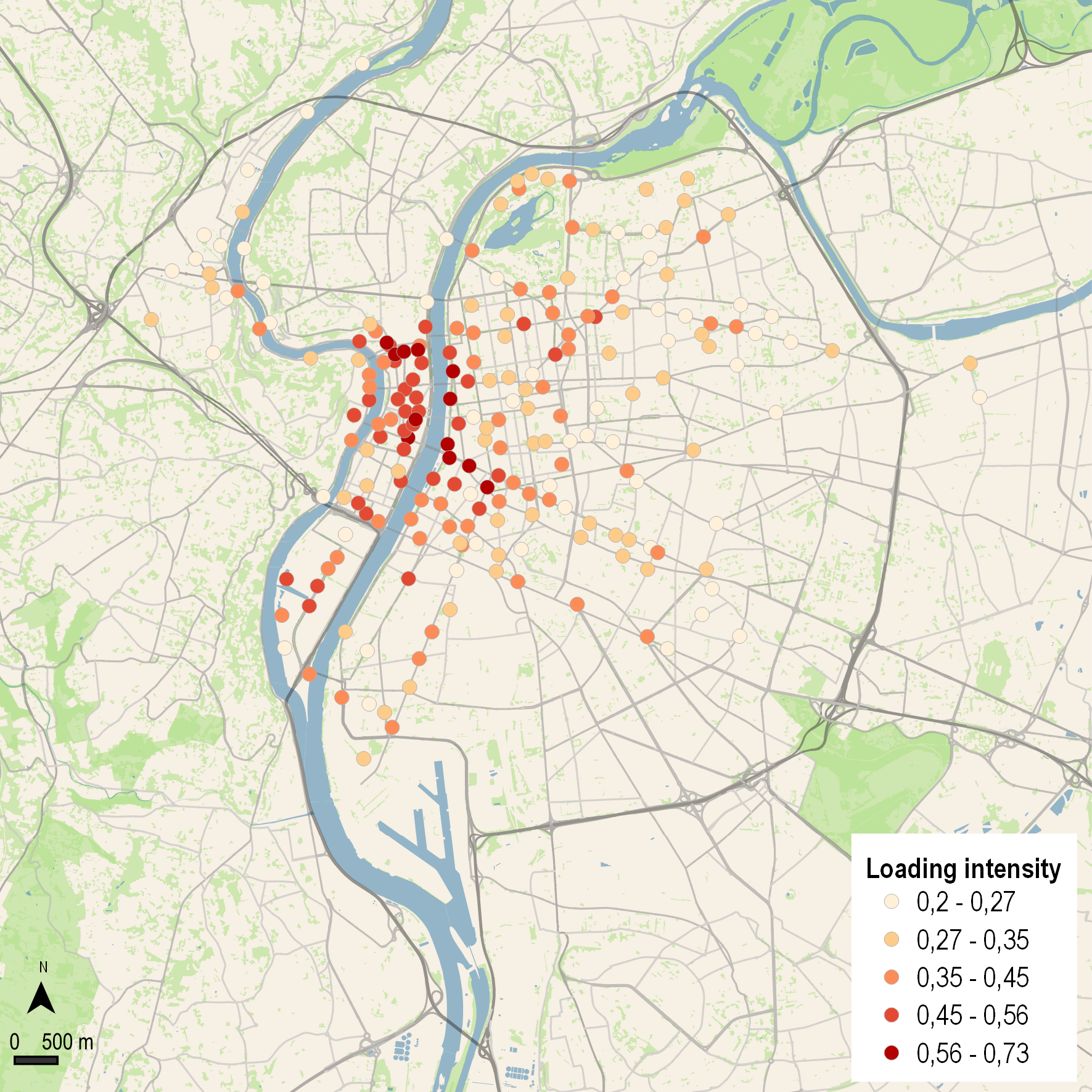}
    
    \vspace{.3cm}
    \includegraphics[width=.8\textwidth]{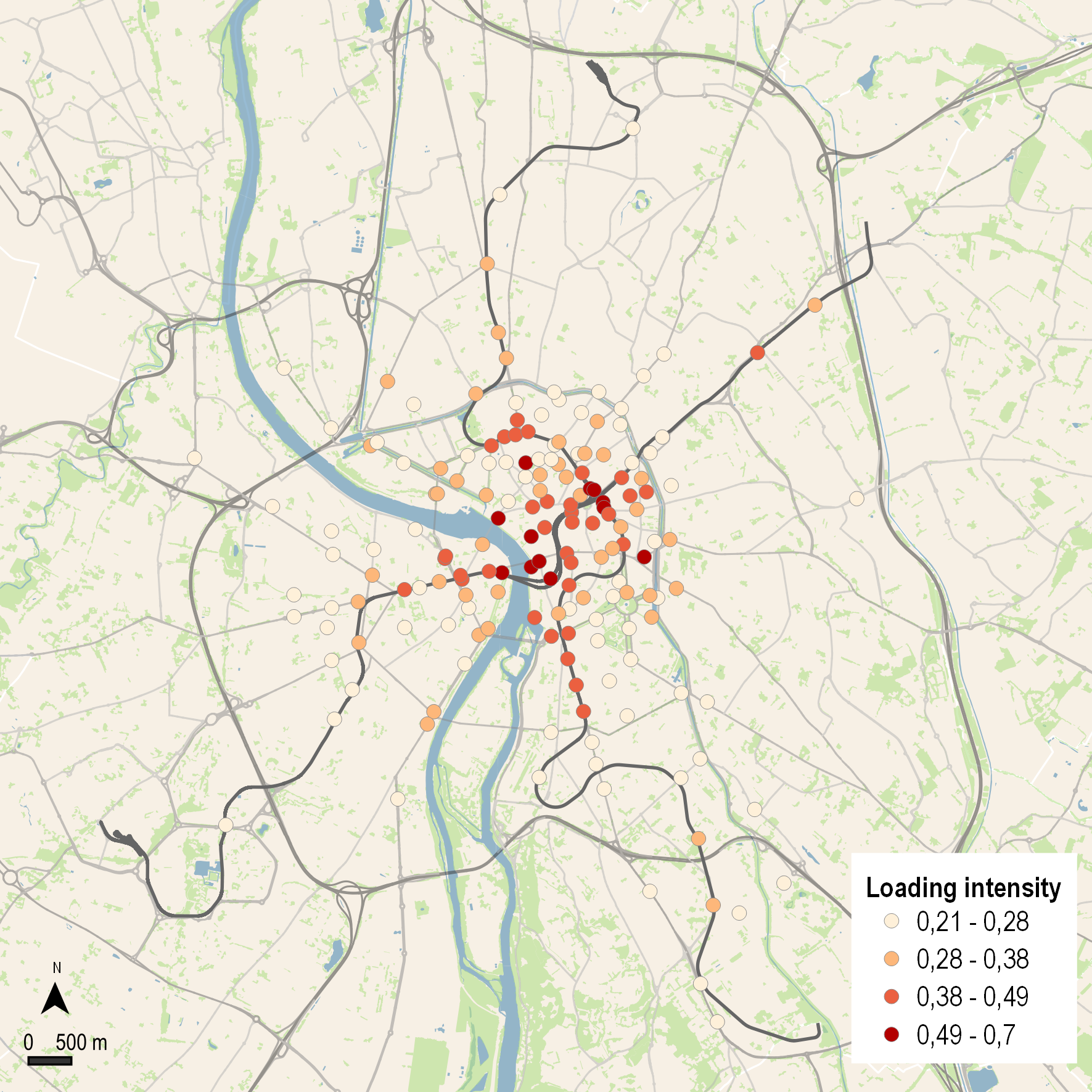}
    \subcaption{$2^{nd}$ factor loading map for Lyon (top) and $3^{rd}$ factor loading map for Toulouse (bottom)}
\end{subfigure}
\caption{The EFA factors representing the late-evening and night-time BSS usage in Lyon and Toulouse.}
\end{figure}
%%%%%%%%%
The daily representation of the scores highlights that this factor is predominant on Saturdays, while loading maps pinpoint bike stations close to areas with high concentration of bars, restaurants and other leisure activities (e.g., Hotel de Ville, Bellecour, Berges du Rhone, and La Guillotiere in Lyon, Place Saint Georges, Place du Capitole, Place Esquirol and Place Saint-Pierre in Toulouse). Our results for Lyon, for example, are consistent with those of Nicolas Chausson, who studied urban nighttime centralities \cite{chausson2019}. Interestingly, the curfew imposed between 9PM and 6AM starting from mid-October is clearly visible on the hourly representations for both towns. Particularly for these last two weeks of October, we observe a peak of traffic between 8PM and 9PM, right before the curfew, probably due to people using bikes to go back home.     

\subsubsection{The parks and stadiums}
%%%%%%%%%%%%%
\begin{figure}[H]
\centering
\begin{subfigure}[b]{0.55\textwidth}
    \centering
    \includegraphics[width=.49\textwidth]{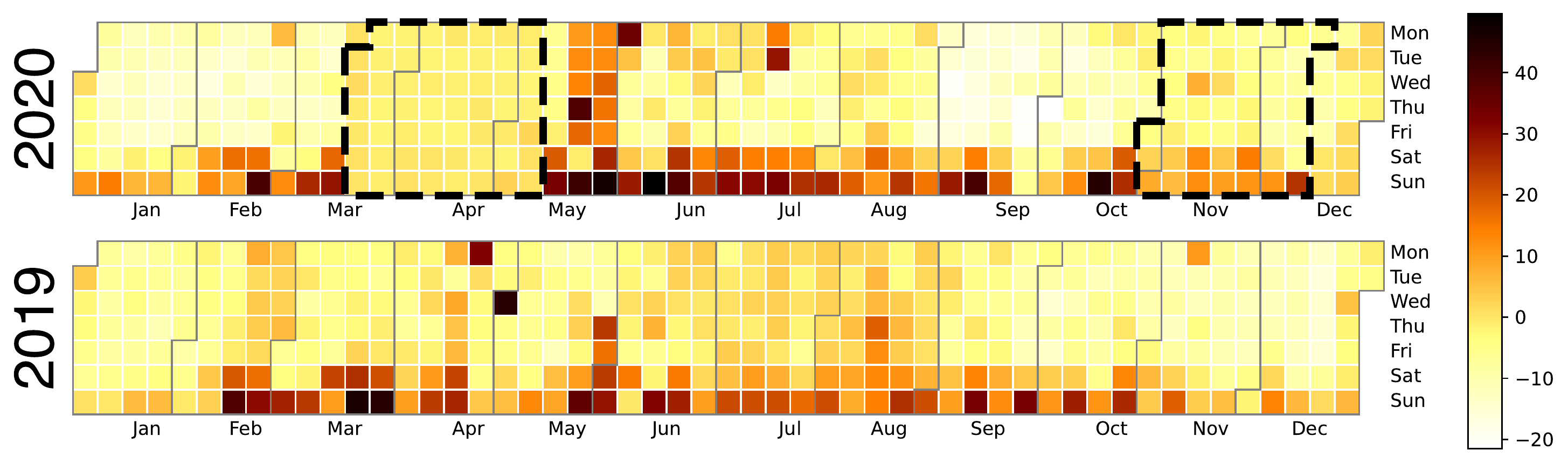}
    \includegraphics[width=.49\textwidth]{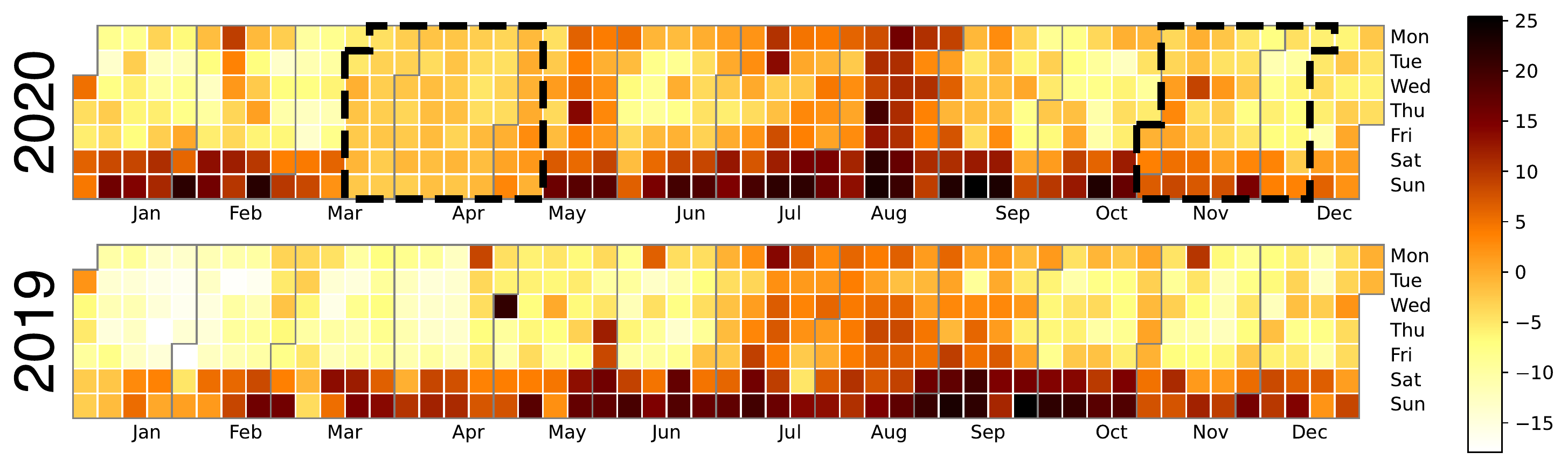}
    \includegraphics[width=.49\textwidth]{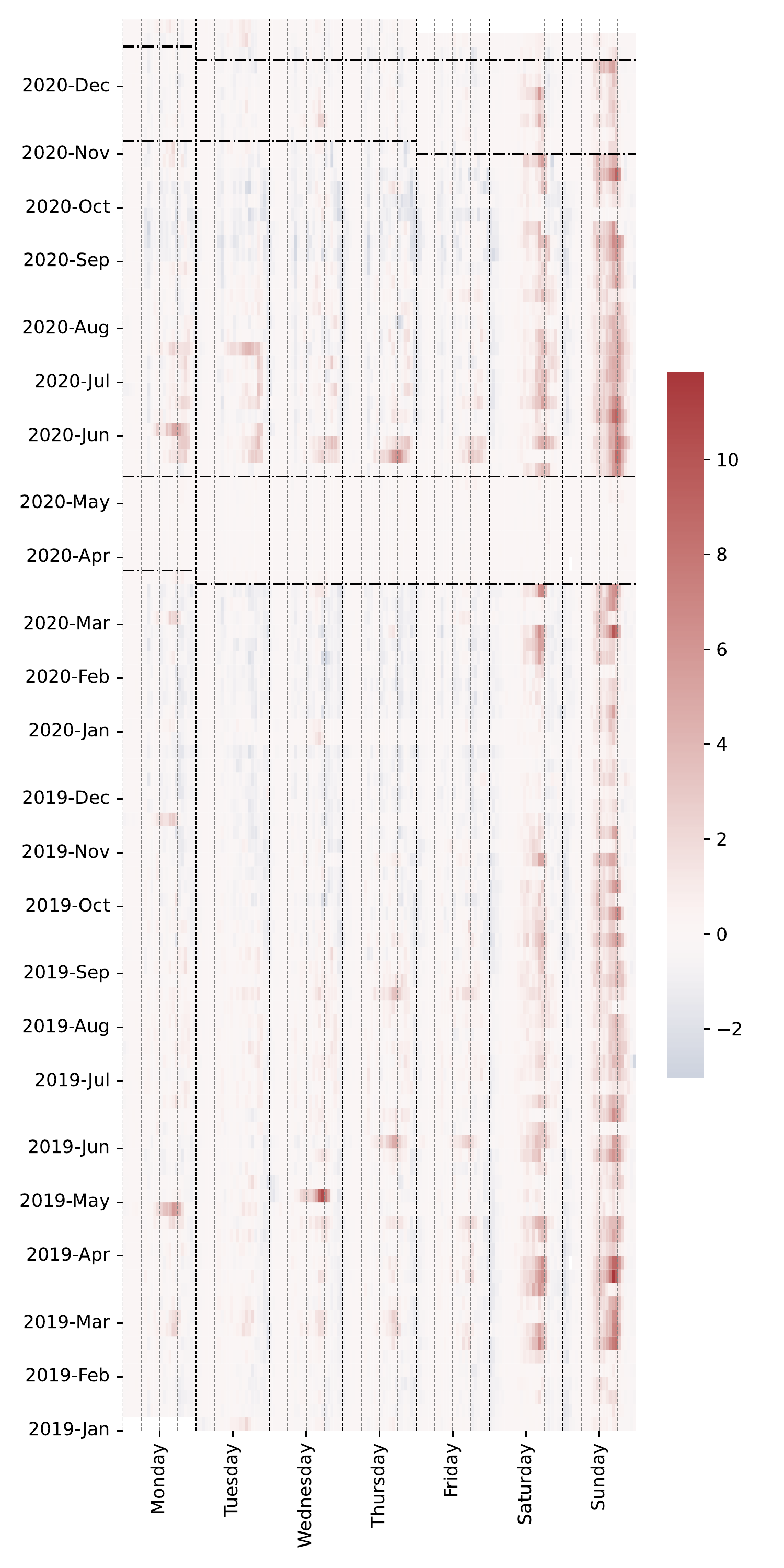}
    \includegraphics[width=.49\textwidth]{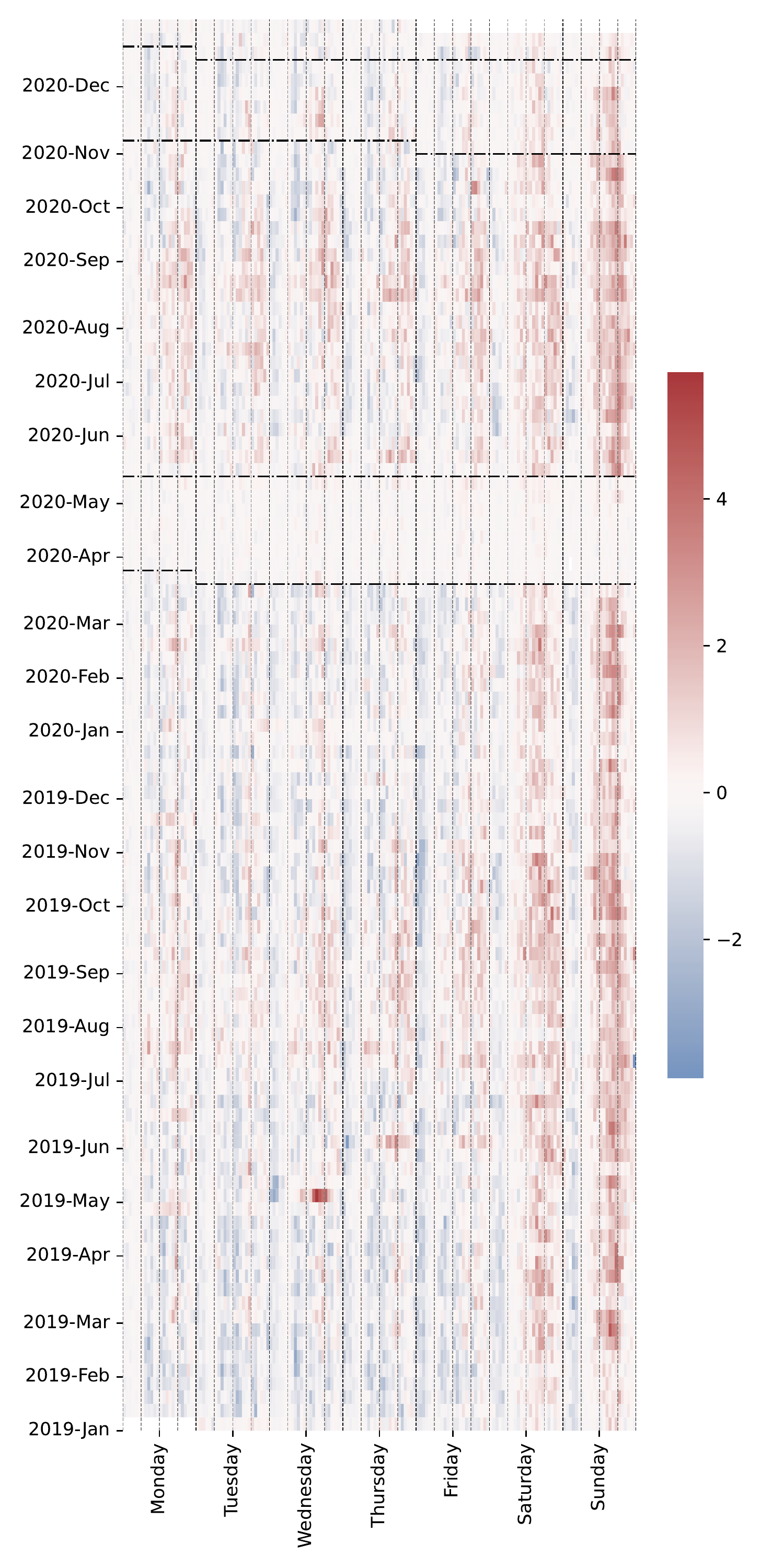}
    \subcaption{Daily and hourly representation of the $6^{th}$ factor scores for Lyon (left) and $5^{th}$ factor scores for Toulouse (right)}
\end{subfigure}
\hspace{1cm}
\begin{subfigure}[b]{0.35\textwidth}
    \centering
    \includegraphics[width=.8\textwidth]{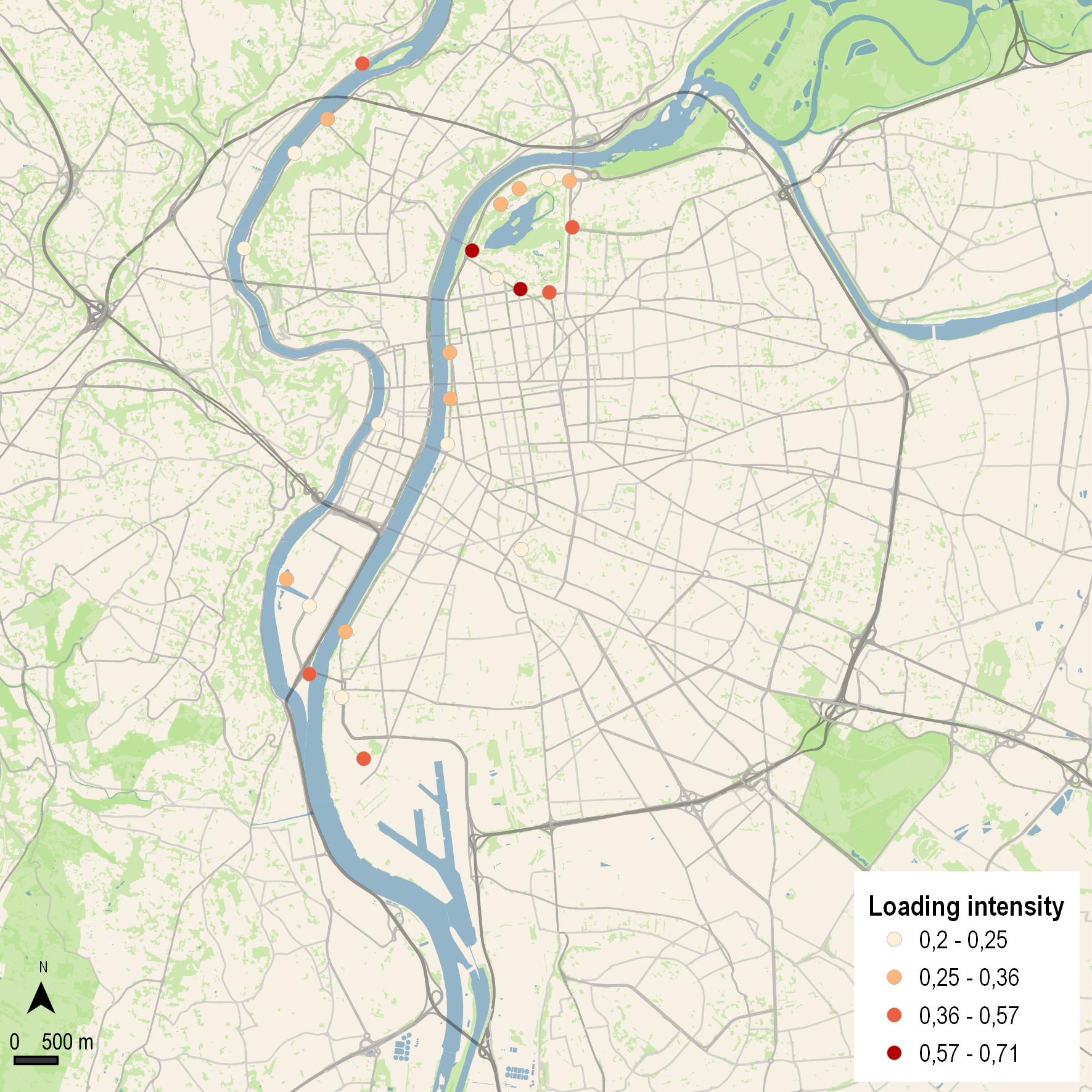}
    
    \vspace{.3cm}
    \includegraphics[width=.8\textwidth]{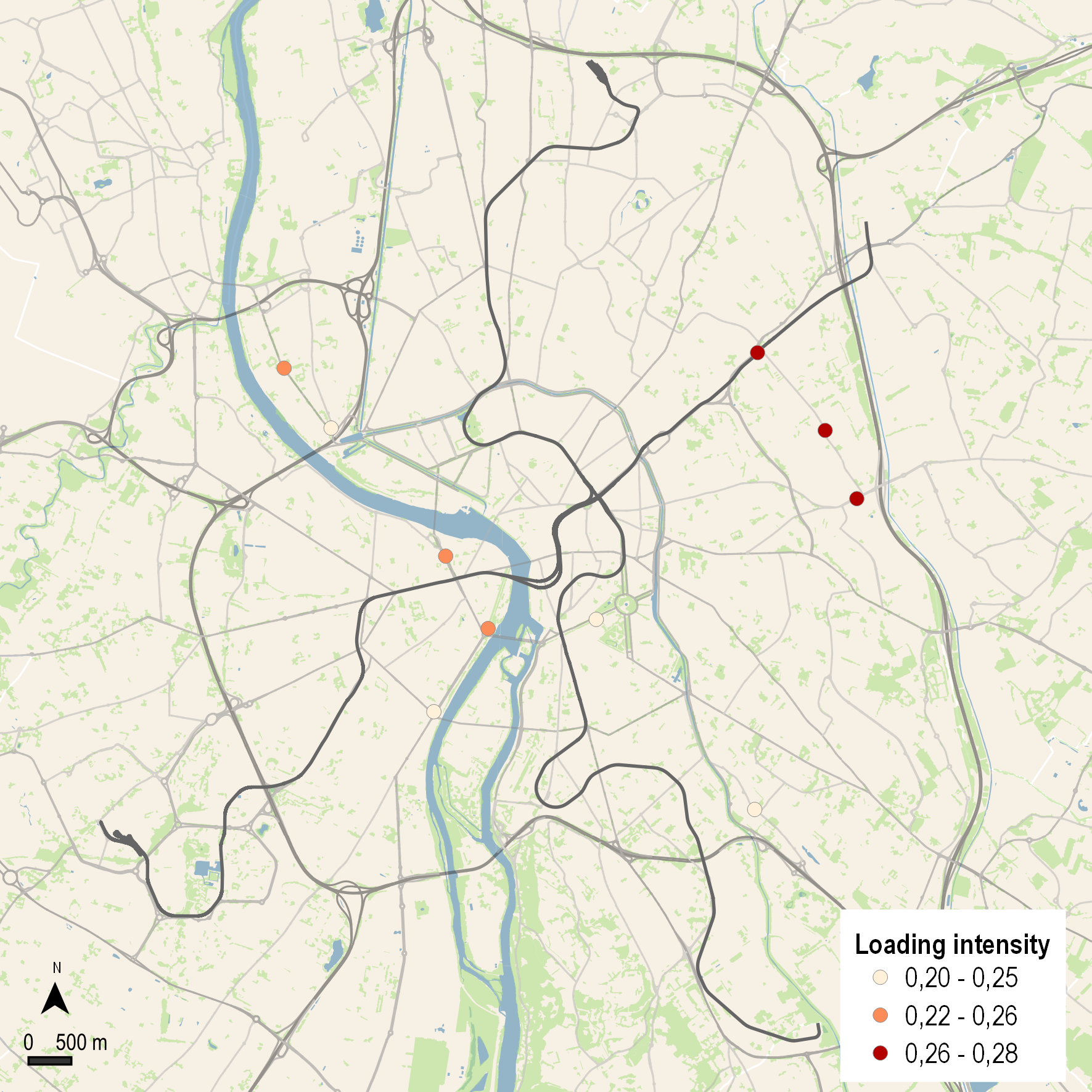}
    \subcaption{$6^{th}$ factor loading map for Lyon (top) and $5^{th}$ factor loading map for Toulouse (bottom)}
\end{subfigure}
\caption{The EFA factors relative to parks and stadiums BSS in Lyon and Toulouse. In the loading maps, only stations with loadings greater than .2 are plotted.}
\label{fig:parks_gardens}
\end{figure}
%%%%%%%%%%%%%%
The $6^{th}$ factor in Lyon and the $5^{th}$ in Toulouse are representative of BSS usage near parks and stadiums (see \Cref{fig:parks_gardens}): Ile Barbe, Saint-Rambert, Tête d'Or, Mulatière and Gerlan in Lyon, Sept Deniers, Jardins Raymond VI, Prairie des Filtres, Jardin des Plantes and les Argoulets in Toulouse. Scores are high during weekends. It can be noted that on the first weekend after the end of the first lockdown (16 and 17 May 2020), bike traffic gets back to pre-pandemic levels near parks and, in an unusual way, remains high during the following two weeks (particularly in Lyon), even though this period did not correspond to vacation time and schools had reopened.  

\section{Conclusion}
\label{Conclusion}
In this paper, we investigate the evolution of Bike-Sharing System (BSS) usage for two French cities, Toulouse and Lyon, during the COVID-19 pandemic situation. 

We first compare the evolution of the number of bike rentals for both towns and along the two years 2019 and 2020, with a 10-minutes time granularity. In this analysis, data from all the dock-stations have been aggregated, thus ignoring any spatial considerations. The results show that these time series can be clearly explained as the product of two main effects: the annual tendency and daily evolution. Our multiplicative regression model shows very good fitting scores for both towns, taking into account school-holidays and rain fall. Secondly, we use Exploratory Factor Analysis (EFA) to derive spatio-temporal patterns in BSS usage over the two years. In the data, dock-station spatial granularity is preserved, thus permitting to observe and interpret, via the retrieved latent factors, the dynamics at the level of dock-stations instead of that of the entire network as in the first study. EFA highlights five and six latent variables, globally explaining around 50\% of the variance for Toulouse and Lyon respectively. The additional latent variable available for Lyon shows a clear student-related BSS usage, which is merged with another factor in the case of Toulouse. 

Among the most relevant changes in the hourly evolution of BSS usage between 2019 (without COVID) and 2020 (with COVID), the global network level analysis shows a proportional increase of BSS use during late morning and early afternoon for working days, in both towns. Nevertheless, the four daily peaks remain located at the same time in 2019 and 2020 in both towns. Concerning BSS traffic volume, morning peaks appear as the most impacted by the pandemic, particularly in Toulouse. The BSS traffic profile in Lyon is definitely more than resilient, it displays a boost of activity between the two lockdowns (from mid-May to end of October), which Toulouse does not. Road traffic and Public-Transport have been much more impacted than BSS. None of them appears to get back to pre-COVID levels. After the first lockdown, with a similar 1-month delay and until the second lockdown, with the exception of summer holidays, road-traffic stabilizes at approximately 85\% of the 2019 values for both towns. The traffic flow of public transport has the same dynamic in both towns, recovering a maximum of 60\% of its value after the first lockdown and 75\% of its value after summer holidays. This is consistent with previous findings in other towns. We found that trip duration significantly increased during the pandemic period for working-days but not for weekend-days. These findings are relevant and original with respect to other analyses performed on other towns. 
Concerning weekend BSS traffic, we also observe that the midday peaks are 15-minute later in Toulouse than in Lyon, in 2019 and 2020. 

Concerning the spatio-temporal analysis performed with EFA, both towns display a morning peak mostly associated with departures from residential areas. This is particularly noticeable in the city of Toulouse, where residential areas appear to be more separate from business ones. An end-of-afternoon peak of BSS traffic is clearly detected by EFA with respect to areas with high density of business and activity centers in the city of Lyon, while it covers a large portion of the whole city center of Toulouse. This BSS usage profile appears to take longer to get back to pre-pandemic levels in Toulouse than in Lyon, even though this phenomenon may be partially explained for Toulouse by student activity (and the related summer holiday break) associated to dock stations included in this pattern. 
For Lyon, a specific factor is associated to student activity, clearly showing the prolonged effect of lockdown, restrictions and holiday breaks which affected student-related bike usages.

Interestingly enough, the EFA analysis clearly identifies a BSS usage pattern in the proximity of parks and stadiums for both cities. After the first lockdown, BSS traffic near these zones appears to be relatively higher in 2020 than in 2019, during weekends and, surprisingly, during the two weeks following the end of the lockdown despite the fact that it was not a vacation period. This could indicate the population's need to reconnect with social and nature activities after lockdown stay-at-home restrictions. 

Our results globally show that BSS appeared as a resilient mobility alternative during the pandemic situation. It probably indicates an increased preference towards cycling in general.  Without surveys, it is however difficult to know which part of the road-traffic decrease can be put down to some new cycling usage, to smart working, or other mobility choices. For this reason, as future research is to shed light on this question, we aim to complete this study by coupling it with an analysis of surveys that are being currently collected in the two analyzed cities. 
The findings of this paper could help transport operators and public authorities better identify and understand cycling behavioral changes in order to respond even more efficiently to an emergency situation, such as the one created by the pandemic, to be prepared for future crises as well as to understand the current opportunities in terms of bicycle use and to set up the conditions for the perpetuation of new cycling practices towards a more resilient and efficient mobility.

\bibliographystyle{alpha}
\bibliography{biblio}

\end{document}